\documentclass{aa}

\usepackage{amsmath}
\usepackage{txfonts}         
\usepackage{natbib}
\usepackage{lscape}          
\usepackage[usenames]{color}
\usepackage{wasysym}         

\definecolor{darkblue}{rgb}{0,0,.4}

\bibpunct{(}{)}{;}{a}{}{,}   

\title{Tidal effects on brown dwarfs: Application to the eclipsing binary 2MASS\,J05352184$-$0546085}

\subtitle{The anomalous temperature reversal in the context of tidal heating}

\author{R. Heller \inst{1} \and B. Jackson \inst{2}  \and R. Barnes \inst{3,4} \and R. Greenberg \inst{5} \and D. Homeier  \inst{6}}

\institute{
Hamburger Sternwarte (Universit\"at Hamburg), Gojenbergsweg 112, 21029 Hamburg, Germany\\ \email{rheller@hs.uni-hamburg.de}
\and
Lunar and Planetary Laboratory, University of Arizona, Tucson, AZ 85721\\
\email{bjackson@lpl.arizona.edu}
\and
University of Washington, Dept. of Astronomy, Seattle, WA 98195
\and
Virtual Planetary Laboratory, NASA\\
\email{rory@astro.washington.edu}
\and
Lunar and Planetary Laboratory, University of Arizona, Tucson, AZ 85721\\
\email{greenber@lpl.arizona.edu}
\and
Institut f\"ur Astrophysik, Georg-August-Universit\"at G\"ottingen, Friedrich-Hund-Platz 1. 37077 G\"ottingen, Germany\\
\email{derek@astro.physik.uni-goettingen.de}
}

\date{Received date / Accepted date}

\abstract
{2MASS\,J05352184$-$0546085 (2M0535$-$05) is the only known eclipsing brown dwarf (BD) binary, and so may serve as a benchmark for models of BD formation and evolution. However, theoretical predictions of the system's properties seem inconsistent with observations: \textit{i.} The more massive (primary) component is observed to be cooler than the less massive (secondary) one.
\textit{ii.} The secondary is more luminous (by $\approx 10^{24}$\,W) than expected. Previous explanations for the temperature reversal have invoked reduced convective efficiency in the structure of the primary, connected to magnetic activity and to surface spots, but these explanations cannot account for the enhanced luminosity of the secondary. Previous studies also considered the possibility that the secondary is younger than the primary.}
{We study the impact of tidal heating to the energy budget of both components to determine if it can account for the observed temperature reversal and the high luminosity of the secondary. We also compare various plausible tidal models to determine a range of predicted properties.}
{We apply two versions of two different, well-known models for tidal interaction, respectively: \textit{i.} the `constant-phase-lag' model and \textit{ii.} the `constant-time-lag' model and incorporate the predicted tidal heating into a model of BD structure. The four models differ in their assumptions about the rotational behavior of the bodies, the system's eccentricity and putative misalignments $\psi$ between the bodies' equatorial planes and the orbital plane of the system.}
{ The contribution of heat from tides in 2M0535$-$05 alone may only be large enough to account for the discrepancies between observation and theory in an unlikely region of the parameter space. The tidal quality factor $Q_{\mathrm{BD}}$ of BDs would have to be $10^{3.5}$ and the secondary needs a spin-orbit misalignment of $\gtrsim 50^\circ$. However, tidal synchronization time scales for 2M0535$-$05 restrict the tidal dissipation function to $\log(Q_{\mathrm{BD}}) \gtrsim 4.5$ and rule out intense tidal heating in 2M0535$-$05. We provide the first constraint on $Q$ for BDs.}
{Tidal heating alone is unlikely to be responsible for the surprising temperature reversal within 2M0535$-$05. But an evolutionary embedment of tidal effects and a coupled treatment with the structural evolution of the BDs is necessary to corroborate or refute this result. The heating could have slowed down the BDs' shrinking and cooling processes after the birth of the system $\approx 1$\,Myr ago, leading to a feedback between tidal inflation and tidal heating. Observations of old BD binaries and measurements of the Rossiter-McLaughlin effect for 2M0535$-$05 can provide further constraints on $Q_{\mathrm{BD}}$.}


\begin{document}

\titlerunning{Tidal effects on brown dwarfs: Application to 2M0535$-$05}

\authorrunning{Heller et al.}

\maketitle

\section{Introduction}
\label{sec:intro}

2MASS\,J05352184$-$0546085 (2M0535$-$05) is a benchmark object for brown dwarf (BD) science since it offers the rare opportunity of independent radius and mass measurements on substellar objects. The observed values constrain evolutionary and structural models \citep{1997MmSAI..68..807D, 1998A&A...337..403B, 2000ARA&A..38..337C, 2002A&A...382..563B, 2007A&A...472L..17C}. 2M0535$-$05 is located in the Orion Nebulae, a star-forming region with an age of 1 ($\pm 0.5$)\,Myr. If both components formed together, as commonly believed, then this system allows for effective temperature ($T_{\mathrm{eff}}$) and luminosity ($L$) measurements of two BDs at the same age.

However, this system is observed to have an unexpected temperature reversal \citep{2006Natur.440..311S}, contravening theoretical simulations: the more massive component (the primary) is the cooler one. From the transit light curve, the ratio of the effective temperatures can be accurately determined to $T_{\mathrm{eff},2}/T_{\mathrm{eff},1} = 1.050 \pm 0.002$ \citep{2009ApJ...697..713M, 2009ApJ...699.1196G}. From spectroscopic measurements then, the absolute values can be constrained. The primary, predicted to have $T_{\mathrm{eff},1} \approx 2\,870$\,K \citep{1998A&A...337..403B}, has an observed value of $\approx 2\,700$\,K, whereas the surface temperature of the secondary, predicted to be $T_{\mathrm{eff},2} \approx 2\,750$\,K, is most compatible with $T_{\mathrm{eff},2} \approx 2\,890$\,K.

One explanation for the temperature discrepancies is suppression of convection due to spots on the surface of the primary. If a portion of a BD's surface is covered by spots, its apparent temperature will be reduced, resulting in an increase in the estimated radius in order for the measured and expected luminosities to agree \citep{2007A&A...472L..17C}. With a spot coverage of 30 - 50\% and a mixing length parameter $\alpha = 1$ most of the mismatches between predicted and observed radii for low-mass stars (LMS) can be explained \citep{2008MmSAI..79..562R}. Observations of spots on both of the 2M0535$-$05 components \citep{2009ApJ...699.1196G}, as inferred from periodic variations in the light curve, and measurements on the H$_{\alpha}$ line of the combined spectrum during the radial velocity maxima \citep{2007ApJ...671L.149R} suggest that enhanced magnetic activity and the accompanying spots on the primary indeed play a key role for its temperature deviation. But even if the spot coverage on the primary serves as an explanation for the primary's reduced $T_{\mathrm{eff}}$, the secondary's luminosity overshoot of $\approx 2.3 \cdot 10^{24}$\,W, as compared to the \citet{1998A&A...337..403B} models, suggests some additional processes may be at work.

The temperature reversal between the primary and secondary may result from a difference between their ages. The secondary could be $\approx 0.5$\,Myr older than the primary, as proposed by \citet{2007ApJ...664.1154S} (see also \citet{1997MmSAI..68..807D}). A difference of 0.5\,Myr could allow the secondary to have converted the necessary amount of gravitational energy into heat\footnote{In contrast to the \citet{1998A&A...337..403B} tracks, the models by \citet{1997MmSAI..68..807D} predict a temperature increase in BDs for the first $\approx 30$\,Myr of their existence.}, which would explain its luminosity excess. But evolutionary models are very uncertain for ages $\lesssim 1$\,Myr \citep{2002A&A...382..563B, 2005AN....326..905W, 2007ApJ...655..541M, 2007IAUS..239..197M} and, in any case, the age determination and physical natures of these very young objects is subject to debate \citep{2008Natur.453.1079S, 2009IAUS..258..161S}. Furthermore, the mutual capture of BDs and LMS into binary systems after each component formed independently is probably too infrequent to account for the large number of eclipsing LMS binaries with either temperature reversals or inflated radii \citep{2001A&A...366..965G, 2007JSARA...1....7C, 2008MmSAI..79..562R, 2009NewA...14..496C, 2009ApJ...691.1400M}.

Here, we consider the role that tidal heating may play in determining the temperatures of the BDs. In Table \ref{tab:parameters} we show the parameters of 2M0535$-$05 necessary for our calculations. The computed energy rates will add to the luminosity of the BDs in some way (Sect. \ref{sub:converting}) and will contribute to a temperature deviation compared to the case without a perturbing body (Sect. \ref{sec:results}). All these energy rates must be seen in the context of the luminosities of the BDs: $L_1 \approx 8.9 \cdot 10^{24}$\,W (luminosity of the primary) and $L_2 \approx 6.6 \cdot 10^{24}$\,W (luminosity of the secondary). At a distance $a$ to the primary component, its luminosity is distributed onto a sphere with area $4\,\pi\,a^2$. The secondary has an effective -- i.e. a 2D-projected -- area of $\pi\,R_2^2$. With $F_{1,\mathrm{a}}$ as the flux of the primary at distance $a$, the irradiation from the primary onto the secondary $L_{1 \rightarrow 2}$ is thus given by

\begin{equation}\label{equ:lum_irrad}
L_{1 \rightarrow 2} = \pi \ R_2^2 \ F_{1,\mathrm{a}} = \pi \ R_2^2 \ \frac{L_1}{4 \ \pi \ a^2} = L_1 \frac{R_2^2}{4 \ a^2} .
\end{equation}

\noindent
Using that equation, we calculate the mutual irradiation of the BDs: $L_{1 \rightarrow 2} \approx 8.5 \cdot 10^{21}$\,W and $L_{2 \rightarrow 1} \approx 1.0 \cdot 10^{22}$\,W. These energy rates are two and three orders of magnitude lower, respectively, than the observed luminosity discrepancy. Hence, we assume that mutual irradiation can be ignored. This simplification is in contrast to the cases of the potentially inflated transiting extrasolar planets WASP-4b, WASP-6b, WASP-12b, and TrES-4, where stellar irradiation \citep{2009arXiv0910.4394I} dominates tidal heating by several magnitudes.

Various tidal models haven been used to calculate tidal heating in exoplanets \citep{2001ApJ...548..466B, 2008MNRAS.391..237J, 2008ApJ...681.1631J, 2009ApJ...695.1006B}, which may in fact be responsible for previous discrepancies between interior models and radii of transiting exoplanets \citep{2008MNRAS.391..237J, 2008ApJ...681.1631J, 2009ApJ...700.1921I}. This success in exoplanets motivates our investigation into BDs. While many different tidal models are available, there is no consensus as to which is the best. For this reason, we apply a potpourri of well-established models to the case of 2M0535$-$05 in order to compare the different results. As we show, tidal heating may account for the temperature reversal and it may have a profound effect on the longer-term thermal evolution of the system.

The coincidence of $P_{\mathrm{orb}} / P_1 \approx 2.9698 \approx 3$, with $P_{\mathrm{orb}}$ as the orbital and $P_1$ primary's rotation period, has been noted before but we assume no resonance between the primary's rotation and the orbit for our calculations. These resonances typically occur in systems with rigid bodies where a fixed deformation of at least one body persists, such as in the Sun-Mercury configuration with Mercury trapped in a 3/2 spin-orbit resonance. We assume that, in the context of tides, BDs may rather be treated as fluids and the shape of the body is not fixed.

With this paper, we present the first investigation of tidal interaction between BDs. In Sect. \ref{sec:tid_mod} we introduce four models for tidal interaction and discuss how we convert the computed energy rates into an increase in effective temperature. Sect. \ref{sec:results} is devoted to the results of our calculations, while we deal with the observational implications in Sect. \ref{sec:discussion}. We end with conclusions about tidal heating in 2M0535$-$05, and in BDs in general, in Sect. \ref{sec:conclusions}.

\section{Tidal Models}
\label{sec:tid_mod}

Two qualitatively different models of tidal dissipation and evolution have been developed over the last century: The `constant-phase-lag' \citep[][Wis08 and FM08 in the following]{1966Icar....5..375G, 2008Icar..193..637W, 2008CeMDA.101..171F}, and the `constant-time-lag' model \citep[][Hut81 in the following]{1981A&A....99..126H}. In the former model, the forces acting on the deformed body are described by a superposition of a static equilibrium potential and a disturbing potential (FM08). The latter model assumes the time between the passage of the perturbing body overhead and the passage of the tidal bulge is constant. Although both models have been used extensively, it is not clear which model provides a more accurate description of the effects of tides, so we apply formulations of both models.

In the `constant-phase-lag' model of FM08, quantitative expressions have been developed to second order in eccentricity $e$ while the others include also higher orders. Higher and higher order expansions require assumptions about the dependence of a body's tidal response to an increasing number of tidal frequencies, which involves considerable uncertainty. Therefore higher order expansions do not necessarily provide more accuracy \citep[FM08;][]{2009ApJ...698L..42G}. In the `constant-phase-lag' model of Wis08, expressions in $e$ are developed to $8^{\mathrm{th}}$ order. The `constant-time-lag' model of Hut81 does not include possible obliquities, while an enhanced version of that model by \citet{2007A&A...462L...5L} (Lev07) does.

Tidal dissipation in BDs has not been observed or even considered previously, and hence, neither model should take precedence when calculating their tidal dissipation, especially since neither tidal model is definitive \citep{2009ApJ...698L..42G}. As our investigation is the first to consider tidal effects on BDs, we will employ several applicable, previously published models to 2M0535$-$05. By surveying a range of plausible models and internal properties, usually encapsulated in the `tidal dissipation function' $Q$ \citep{1966Icar....5..375G}, we may actually be able to determine which model is more applicable to the case of BDs -- assuming, of course, that tidal dissipation contributes crucially to the observed temperature inversion.

\renewcommand{\arraystretch}{1.3}
\begin{table}[h]
  \centering
  \caption{Orbital and physical parameters of 2M0535$-$05}
  \label{tab:parameters}

    \begin{tabular}{l|c}

    \hline

    \hline

    \hline \hline

    \multicolumn{1}{c}{\textsc{property} }& \textsc{observed value}\\
    \hline

    $a$, semi-major axis$^1$ & 0.0407 $\pm 0.0008$\,AU  \\

    $e$, eccentricity$^1$ & 0.3216 $\pm 0.0019$ \\

    $P_{\mathrm{orb}}$, orbital period$^1$ & 9.779556 $\pm 0.000019$\,d \\

    $i$, orbital inclination to the line of sight$^1$ & 88.49 $\pm 0.06^\circ$ \\

    age$^1$ & 1 $\pm 0.5$\,Myr \\

    $T_{\mathrm{eff},1}$, primary effective temperature$^1$ & 2\,715 $\pm 100$\,K \\

    $T_{\mathrm{eff},2}/T_{\mathrm{eff},1}$, effective temperature ratio$^1$ & 1.050 $\pm 0.002$ \\

    $M_{1}$, primary mass$^1$ & 0.0572 $\pm 0.0033$\,$M_{\odot}$ \\

    $M_{2}$, secondary mass$^1$ & 0.0366 $\pm 0.0022$\,$M_{\odot}$ \\

    $R_{1}$, primary radius$^1$ & 0.690 $\pm 0.011$\,$R_{\odot}$ \\

    $R_{2}$, secondary radius$^1$ & 0.540 $\pm 0.009$\,$R_{\odot}$ \\

    $L_{1}$, primary luminosity$^3$ & $8.9 \cdot 10^{24}$ $\pm 3 \cdot 10^{24}$\,W \\

    $L_{2}$, secondary luminosity$^3$ & $6.6 \cdot 10^{24}$ $\pm 2 \cdot 10^{24} $\,W \\

    $P_{1}$, rotational period of the primary$^1$  & 3.293 $\pm 0.001$\,d \\

    $P_{2}$, rotational period of the secondary$^1$ & 14.05 $\pm 0.05$\,d \\

    $ \bar{T}_{\mathrm{eff},1}$, modeled $T_{\mathrm{eff}}$ for the primary$^2$ & 2\,850\,K\\

    $ \bar{T}_{\mathrm{eff},2}$, modeled $T_{\mathrm{eff}}$ for the secondary$^2$ & 2\,700\,K\\

    $ \bar{R}_1$, modeled radius for the primary$^2$ & $0.626 \, R_{\odot}$\\

    $ \bar{R}_2$, modeled radius for the secondary$^2$ & $0.44 \, R_{\odot}$\\

    \hline
    \multicolumn{2}{l}{}\\
    \multicolumn{2}{l}{$^1$ \citet{2009ApJ...699.1196G}, $^2$ \citet{1998A&A...337..403B},}\\
    \multicolumn{2}{l}{$^3$ assuming an uncertainty of 200\,K in $T_{\mathrm{eff},1}$ and $T_{\mathrm{eff},2}$}

  \end{tabular}
\end{table}

\subsection{Constant phase lag}
\label{sub:const_ang_lag}

\subsubsection{Tidal model \#1}
\label{subsub:tid_model_1}

The potential of the perturbed body can be treated as the superposition of periodic contributions of tidal frequencies at different phase lags and the expression for the potential can be expanded to first order in those lags (FM08). Those phase lags $\varepsilon_{k,i \ | \ k = 0, 1, 2, 5, 8, 9}$ of the $i^{\mathrm{th}}$ body that we will need for our equations are given by

\begin{align}\label{equ:epsilon}
\nonumber
Q_i \ \varepsilon_{0,i} & = \Sigma(2 \Omega_i - 2 n)\\
\nonumber
Q_i \ \varepsilon_{1,i} & = \Sigma(2 \Omega_i - 3 n)\\
\nonumber
Q_i \ \varepsilon_{2,i} & = \Sigma(2 \Omega_i - n)\\
\nonumber
Q_i \ \varepsilon_{5,i} & = \Sigma(n)\\
\nonumber
Q_i \ \varepsilon_{8,i} & = \Sigma(\Omega_i - 2 n)\\
Q_i \ \varepsilon_{9,i} & = \Sigma(\Omega_i) \hspace{1.5cm} i \in \{1,2\} ,\\
\nonumber
\end{align}

\noindent
where $\Sigma(x)$ is the algebraic sign of $x$, thus $\Sigma(x)~=~+1~\vee~-1$, $n = 2 \pi / P_{\mathrm{orb}}$ is the orbital frequency and $\Omega_i = 2 \pi / P_i$ are the rotational frequencies of the primary ($i = 1$) and secondary ($i = 2$), $P_i$ being their rotational periods. The tidal frequencies are functions of the tidal quality factor $Q$ of the deformed object, which parametrizes the object's tidal response to the perturber. It is defined as

\begin{equation}
Q^{-1} = \frac{1}{2 \pi E_0} \int_0^{P_{\mathrm{orb}}} \mathrm{d}t \left( - \frac{\mathrm{d}E}{\mathrm{d}t} \right) ,
\end{equation}

\noindent
where $E_0$ is the maximum energy stored in the tidal distortion and the integral over the energy dissipation rate $-$d$E$/d$t$ is the energy lost during one orbital cycle \citep{1966Icar....5..375G}. Although \citet{2004ApJ...610..477O} conclude that tidal dissipation rates of giant planets are not adequately represented by a constant $Q$-value, many parameterized tidal models rely on this quantity. Measurements of the heat flux from Jupiter's moon Io during the fly-by of the Voyager 1 spacecraft, combined with a specific model of the history of the orbital resonance, allowed for an estimate for the quality factor $Q_{\jupiter}$ of Jupiter to be $2 \cdot 10^{5} < Q_{\jupiter} < 2 \cdot 10^{6}$ \citep{1979Natur.279..767Y} while \citet{2001AJ....122.2734A} used historical changes in Io's orbit to infer that $Q_{\jupiter}$ is around $10^{5.3}$. However, \citet{2008DPS....40.0403G} pointed out that $Q = \infty$ is not ruled out \citep[see also][]{1980LPI....11..871P, 1993ApJ...406..266I}. Tides raised by Neptune on its moons help to constrain the planet's quality factor to $10^{3.95} < Q_{\neptune} < 10^{4.56}$ \citep{2008Icar..193..267Z}. For M dwarfs, $Q_{\mathrm{dM}}$ is assumed to be of order $10^5$, whereas for rigid bodies like Earth $20 \lesssim Q \lesssim 500$ \cite[][and references therein]{2001GeoJI.144..471R, 2004ApJ...614..955M}. For BDs, however, $Q$ is even more uncertain, thus we will handle it as a free parameter in our procedures.

FM08 allows for the tidal amplitude to be different from what it would be if the tide-raising body were fixed in space. This concern is met by the dynamical Love number $k_\mathrm{d}$ under the assumption that the tidally disturbed body had infinite time to respond. Without better knowledge of a body's response to tides, we assume the dynamical Love number is the same as the potential Love number of degree 2, $k_2$. For the gas planets of the solar system, this number has been calculated by \citet{1977Icar...32..443G}. BDs may rather be treated as polytropes of order $\mathfrak{n} = 3/2$ (I. Baraffe, private communication). We infer the Love number from the relation $k_\mathrm{2} = 2 k_\mathrm{aps}$ \citep{2002ApJ...573..829M} and use the tables of apsidal motion constants $k_\mathrm{aps}$ given in \citet{1955MNRAS.115..101B}. These authors provide numerical calculations for $k_{\mathrm{aps}}$ for a polytrope of $\mathfrak{n} = 3/2$. We find $k_\mathrm{aps} = 0.143$ and thus $k_\mathrm{d} \equiv k_\mathrm{2} = 0.286$. This places $k_2$ for BDs well in the regime spanned by the gas giants of the solar system: Jupiter ($k_2 = 0.379$), Saturn ($k_2 = 0.341$), Uranus ($k_2 = 0.104$) and Neptune ($k_2 = 0.127$) \citep{1977Icar...32..443G}.

Before we proceed to the equations for the tidal heating rates, we sum up those for the orbital evolution of the semi-major axis $a$, the eccentricity $e$ and the putative obliquity $\psi$. The latter parameter is the angle between the equatorial plane of one of the two bodies in a binary system and the orbital plane \citep{2005ApJ...631.1215W}, frequently referred to as spin-orbit misalignment. We use Eqs. (56), (60) and (61) from FM08 but our equations for a binary system with comparable masses need slight modifications since both constituents contribute significantly to the evolution of $a$ and $e$. We add both the terms for the secondary being the perturber of the primary ($i = 1$, $j = 2$) and vice versa, since only spin-orbit coupling is relevant, whereas spin-spin interaction can be neglected. This results in

\begin{align}\label{equ:a_FM}
\nonumber
  \frac{\mathrm{d}a}{\mathrm{d}t} = \sum_{ \substack{i \, = \, 1,2 \\ i \, \neq \, j} } \frac{3 k_{\mathrm{d},i} M_j R_i^5 n}{4 M_i a^4} \ ( \ &4 \varepsilon_{0,i} + e^2 [ - 20 \varepsilon_{0,i} + \frac{147}{2} \varepsilon_{1,i} + \frac{1}{2} \varepsilon_{2,i}\\
&- 3 \varepsilon_{5,i}] - 4 S_i^2 [\varepsilon_{0,i} - \varepsilon_{8,i}] \ ) ,
\end{align}

\begin{equation}\label{equ:e_FM}
  \frac{\mathrm{d}e}{\mathrm{d}t} = - \sum_{ \substack{i \, = \, 1,2 \\ i \, \neq \, j} } \frac{3 e k_{\mathrm{d},i} M_j R_i^5 n}{8 M_i a^5} \left( 2 \varepsilon_{0,i} - \frac{49}{2} \varepsilon_{1,i} + \frac{1}{2} \varepsilon_{2,i} + 3 \varepsilon_{5,i} \right) ,
\end{equation}

\begin{equation}\label{equ:psi_FM}
  \frac{\mathrm{d}\psi_i}{\mathrm{d}t} = \frac{3 k_{\mathrm{d},i} M_j R_i^5 n}{4 M_i a^5} \ S_i \ \left( - \varepsilon_{0,i} + \varepsilon_{8,i} + - \varepsilon_{9,i} \right) ,
\end{equation}

\noindent
where $k_{\mathrm{d},i}$ is the dynamical Love number, $M_i$ the mass and $R_i$ the radius of the deformed BD, $S_i \coloneqq \sin(\psi_i)$, with $\psi_i$ as the obliquity of the perturbed body, and $\varepsilon_{k,i \ | \ k = 0, 1, 2, 5, 8, 9}$ are the tidal phase lags, given in Eq. (\ref{equ:epsilon}).

The total energy that is dissipated within the perturbed body, its tidal energy rate, can be determined by summing the work done by tidal torques (Eqs. (48) and (49) in FM08). The change in orbital energy of the $i^{\mathrm{th}}$ body due to the $j^{\mathrm{th}}$ body is given by

\begin{align}\label{equ:E_orb_mod1}
\nonumber
\dot{E}_{\mathrm{orb},i}^{\mathrm{\#1}} = \
\underbrace{ \frac{3 k_{\mathrm{d},i} G M_j^2 R_i^5}{8 a^6} }_p
\ n \ ( \ &4 \varepsilon_{0,i} + e^2 [-20 \varepsilon_{0,i} + \frac{147}{2} \varepsilon_{1,i} + \frac{1}{2} \varepsilon_{2,i}\\
&- 3 \varepsilon_{5,i} ] - 4 S_i^2 \ [\varepsilon_{0,i} - \varepsilon_{8,i}] \ )
\end{align}

\noindent
and the change in rotational energy is deduced to be

\begin{align}\label{equ:E_rot_mod1}
\nonumber
\dot{E}_{\mathrm{rot},i}^{\mathrm{\#1}} = \ - \frac{3 k_{\mathrm{d},i} G M_j^2 R_i^5}{8 a^6} \ \Omega_i \ (& \ 4 \varepsilon_{0,i} + e^2 [-20 \varepsilon_{0,i} + 49 \varepsilon_{1,i} + \varepsilon_{2,i}]\\
& + 2 S_i^2 \ [- 2 \varepsilon_{0,i} + \varepsilon_{8,i} + \varepsilon_{9,i}] \ ) ,
\end{align}

\noindent
where $G$ is Newton's gravitational constant. The total energy released inside the body then is

\begin{equation}\label{equ:E_tid_mod1}
\dot{E}_{\mathrm{tid},i}^{\mathrm{\#1}} = - \ (\dot{E}_{\mathrm{orb},i}^{\mathrm{\#1}} + \dot{E}_{\mathrm{rot},i}^{\mathrm{\#1}}) > 0 .
\end{equation}

\noindent
The greater-than sign in this equation is true, since either $\Omega_i < n$ and orbital energy is converted into rotational energy, or $\Omega_i > n$ and the body is decelerated by a transfer of rotational energy into orbital energy. In both cases, the dynamical energy of the system is released within the distorted body. For $\Omega_i = 0$, e.g., Eqs. (\ref{equ:E_orb_mod1}) and (\ref{equ:E_rot_mod1}) yield $\dot{E}_{\mathrm{orb},i}^{\mathrm{\#1}} = -p \cdot (4+57e^2+4S_i^2) / Q_i$ and $\dot{E}_{\mathrm{rot},i}^{\mathrm{\#1}} = 0$.

The approach for the calculation of tidal energy rates with tidal model \#1 depends on processes due to non-synchronous rotation via $\varepsilon_{k,i} = \varepsilon_{k,i}(\Omega_i,n)$ and includes a putative obliquity $\psi_i$ and terms of $e$ up to the second order. After inserting the orbital and rotational periods for 2M0535$-$05, these equations reduce to

\begin{align}\label{equ:E_tid_mod1_2M}
\nonumber
\dot{E}_{\mathrm{tid,1}}^{\mathrm{\#1}} & =  \frac{3 k_{\mathrm{d},1} G M_2^2 R_1^5 }{8 Q_1 a^6} \ \left( \ [ 4 + 30 e^2 ] \Omega_1 - [4 + 51 e^2] n \ \right) ,\\
\dot{E}_{\mathrm{tid,2}}^{\mathrm{\#1}} & =  \frac{3 k_{\mathrm{d},2} G M_1^2 R_2^5 }{8 Q_2 a^6} \ \left( \ [ 4 + 56 e^2 ] n + [2 S_2^2 - 4 - 28 e^2] \Omega_2 \ \right) .
\end{align}

\noindent
Interestingly, for these particular values of $\Omega_1$, $\Omega_2$ and $n$, the $S_1$-terms for $\dot{E}_{\mathrm{tid},1}^{\mathrm{\#1}}$ cancel each other, so that it is not a function of $\psi_1$, whereas $\dot{E}_{\mathrm{tid},2}^{\mathrm{\#1}}$ does depend on $\psi_2$.

\subsubsection{Tidal model \#2}

The model of Wis08 includes terms in eccentricity up to the $8^\mathrm{th}$ order, predicting higher tidal energy rates than for the equations of model \#1. Equations for the evolution of the orbital parameters are not given in Wis08. Furthermore, in his theory the perturbed body is assumed to be synchronously rotating with the orbital period. Since this is not the case for either of the BDs in 2M0535$-$05, the following equations will only yield lower limits for the tidal heating. The tidal heating rates are given by

\begin{equation}\label{equ:E_tid_mod2}
\dot{E}_{\mathrm{tid},i}^{\mathrm{\#2}} =\frac{21 k_{2,i} G M_j^2 R_i^5 n}{2 Q_i a^6} \zeta_{\mathrm{Wis}}(e,\psi_i)
\end{equation}

\noindent
with

\begin{align}\label{equ:zeta_Wis}
\nonumber
\zeta_{\mathrm{Wis}}(e,\psi_i) = &\frac{2}{7} \frac{f_1^{\mathrm{Hut}}}{\beta^{15}} -  \frac{4}{7}  \frac{f_2^{\mathrm{Hut}}}{\beta^{12}} \ C_i +  \frac{1}{7}  \frac{f_5^{\mathrm{Hut}}}{\beta^9} \left(1+C_i^2\right)\\
& + \frac{3}{14}  \frac{e^2 f_3^{\mathrm{Wis}}}{\beta^{13}} \ S_i^2 \cos(2 \Lambda_i) ,
\end{align}

\noindent
where we used $C_i \coloneqq \cos(\psi_i)$ and

\begin{align}\label{equ:f_Hut_1}
\nonumber
\beta & = \sqrt{1-e^2} ,\\
\nonumber
f_1^{\mathrm{Hut}} & = 1 + \frac{31}{2} e^2 + \frac{255}{8} e^4 + \frac{185}{16} e^6 + \frac{25}{64} e^8 ,\\
\nonumber
f_2^{\mathrm{Hut}} & = 1 + \frac{15}{2} e^2 + \frac{45}{8} e^4 + \frac{5}{16} e^6 ,\\
\nonumber
f_5^{\mathrm{Hut}} & = 1 + 3 e^2 + \frac{3}{8} e^4 ,\\
f_3^{\mathrm{Wis}} & = 1 - \frac{11}{6} e^2 + \frac{2}{3} e^4 + \frac{1}{6} e^6 ,\\
\nonumber
\end{align}

\noindent
following the nomenclature of \citet{1981A&A....99..126H} and \citet{2008Icar..193..637W} as indicated. Furthermore, $k_{2,i}$ is the potential Love number of degree 2 for the $i^{\mathrm{th}}$ component of the binary system and $\Lambda_i$ is a measure of the longitude of the node of the body's equator on the orbit plane with respect to the pericenter of its orbit. In order to estimate the impact of $\Lambda_i$ in the last term in Eq. (\ref{equ:zeta_Wis}), we assume this impact to be as large as possible, $\Lambda_i = 0$, and compare it to the preceding terms. We find that for the case of 2M0535$-$05 the first three terms are of order 1, whereas the term connected to $\Lambda_i$ varies between $10^{-2}$ and $10^{-5}$, depending on $\psi_i$. These irrelevant contributions give us a justification to neglect the unknown values of $\Lambda_i$ in 2M0535$-$05 for our computations, facilitating the comparisons to the other models.

\subsection{Constant time lag}
\label{sub:const_time_lag}

\subsubsection{Tidal model \#3}
\label{subsub:tid_model_3}

Instead of assuming phase lags and superposition of frequency-dependent potentials, the `equilibrium tide' model by \citet{1981A&A....99..126H} invokes a constant time lag $\tau$ between the line joining the centers of the two bodies and the culmination of the tidal bulge on the distorted object. With that assumption, the model of Hut81 is mutually exclusive with the assumption of a fixed angle lag \citep{1966Icar....5..375G}: in general, a fixed time lag and a fixed angle lag result in very different behaviors of the tidal bulge\footnote{If $e = 0$ and $\psi = 0$, then there is a single tidal lag angle $\varepsilon$ and the tidal dissipation funtion can be written as $Q = 1/\varepsilon = 1/(\tau n)$. For the course of an orbit, where the tidal evolution of $n$ is negligible, both $Q$ and $\tau$ can be fixed. However, in a general case where $\tau$ is constant in time, $Q$ will decrease as the orbital semi-major axis decays and $n$ increases. So $Q$ would not be constant.}. As for the case of the `constant-time-lag' model, we first sum up the equations governing the behavior of the orbital evolution. With the purpose of easing a comparison between Hut81's equations (Eqs. (9), (10), and (11) therein) and Eqs. (\ref{equ:a_FM}) - (\ref{equ:psi_FM}) from this paper for the theory of the `constant-phase-lag' model \#1, we transform the former into

\begin{equation}\label{equ:a_Hut}
 \frac{\mathrm{d}a}{\mathrm{d}t} = \sum_{ \substack{i \, = \, 1,2 \\ i \, \neq \, j} } \frac{- 6 k_{\mathrm{aps},i} G M_j R_i^5}{a^7} \ \tau_i \ \left( 1 + \frac{M_j}{M_i} \right) \left( \frac{f_1^{\mathrm{Hut}}}{\beta^{15}} - \frac{f_2^{\mathrm{Hut}}}{\beta^{12}} \frac{\Omega_i}{n} \right) ,
\end{equation}

\begin{equation}\label{equ:e_Hut}
 \frac{\mathrm{d}e}{\mathrm{d}t} = \sum_{ \substack{i \, = \, 1,2 \\ i \, \neq \, j} } \frac{- 27 k_{\mathrm{aps},i} G M_j R_i^5 e}{a^8} \ \tau_i \ \left( 1 + \frac{M_j}{M_i} \right) \left( \frac{f_3^{\mathrm{Hut}}}{\beta^{13}} - \frac{11}{18}\frac{f_4^{\mathrm{Hut}}}{\beta^{10}} \frac{\Omega_i}{n} \right) ,
\end{equation}

\begin{align}\label{equ:psi_Hut}
\nonumber
 \frac{\mathrm{d}\psi_i}{\mathrm{d}t} = &\frac{- 3 k_{\mathrm{aps},i} G M_j^2 R_i^3 \psi_i}{M_i a^6 r_{\mathrm{g},i}^2} \ \tau_i\\
& \times \left( \frac{f_2^{\mathrm{Hut}}}{\beta^{12}} \frac{n}{\Omega_i}
- \frac{f_5^{\mathrm{Hut}}}{2 \beta^9} \left[ 1- \frac{r_{\mathrm{g},i}^2}{\beta} \frac{M_i + M_j}{M_j} \left( \frac{R_i}{a} \right)^2 \frac{\Omega_i}{n} \right] \right) ,
\end{align}

\noindent
with $k_{\mathrm{aps},i}$ as the apsidal motion constant of the perturbed body (see Sect. \ref{subsub:tid_model_1}), $r_{\mathrm{g},i}^2$ as the radius of gyration of the $i^{\mathrm{th}}$ body, which is defined by the body's moment of inertia $I_i = M_i r_{\mathrm{g},i}^2 R_i^2$, and

\begin{align}\label{equ:f_Hut_2}
\nonumber
f_3^{\mathrm{Hut}} & = 1 + \frac{15}{4} e^2 + \frac{15}{8} e^4 + \frac{5}{64} e^6 ,\\
f_4^{\mathrm{Hut}} & = 1 + \frac{3}{2} e^2 + \frac{1}{8} e^4 .
\end{align}

\noindent
Hut81 then calculates the energy dissipation rate within a binary system, caused by the influence of one of the two bodies on the other, as the change in the total energy $E = E_\mathrm{orb} + E_\mathrm{rot}$. Here, $E_\mathrm{orb}$ and $E_\mathrm{rot}$ are the orbital and rotational energies of the body (Eqs. (A28) - (A35) in Hut81). For the tidal heating rates of the $i^{\mathrm{th}}$ constituent within the binary, this yields

\begin{equation}\label{equ:E_tid_mod3}
 \dot{E}_{\mathrm{tid},i}^{\mathrm{\#3}} = \frac{3 k_{\mathrm{aps},i} G M_j^2 R_i^5 n^2}{a^6} \ \tau_i \ \zeta_{\mathrm{Hut}}(e,\Omega_i,n) ,
\end{equation}

\noindent
where

\begin{equation}\label{equ:zeta_Hut}
 \zeta_{\mathrm{Hut}}(e,\Omega_i,n) = \frac{f_1^{\mathrm{Hut}}}{\beta^{15}} - 2 \frac{f_2^{\mathrm{Hut}}}{\beta^{12}} \frac{\Omega_i}{n} + \frac{f_5^{\mathrm{Hut}}}{\beta^{9}} \frac{\Omega_i^2}{n^2} .
\end{equation}

\noindent
Unfortunately, with these equations for the tidal energy rates model \#3 neglects a potential obliquity of the body, which prevents us from a direct comparison with the other tidal models.

\subsubsection{Tidal model \#4}
\label{subsub:tid_Lev}

Lev07 extended Hut81's formula for the tidal energy rate to the case of an object in equilibrium rotation\footnote{Wis08 calls this `asymptotic nonsynchronous rotation'.} and they included possible obliquities \citep[see also][]{1997A&A...318..975N}, though they do not give the equations for the orbital evolution. Lev07's equations are equivalent to

\begin{equation}\label{equ:E_tid_mod4}
 \dot{E}_{\mathrm{tid},i}^{\mathrm{\#4}} = \frac{3 k_{2,i} G M_j^2 R_i^5 n}{Q_{\mathrm{n},i} a^6} \ \zeta_{\mathrm{Lev}}(e,\psi_i) ,
\end{equation}

\noindent
where

\begin{equation}
 \zeta_{\mathrm{Lev}}(e,\psi_i) = \frac{f_1^{\mathrm{Hut}}}{\beta^{15}} - \frac{ ( \ f_2^{\mathrm{Hut}} \ / \ \beta^{12} \ )^{2}}{f_5^{\mathrm{Hut}} \ / \ \beta^{9}} \ \left(1 + \frac{1}{1 - 2/S_i^2}\right)
\end{equation}

\noindent
The `annual tidal quality factor' is given as $Q_{\mathrm{n}}^{-1} = n \ \tau$. Even though Lev07's equations invoke $Q_{\mathrm{n}}$ and their equations resemble those of the models with constant phase lag, their approach still assumes a constant-time-lag. Since Lev07 do not explicitly connect their $Q_{\mathrm{n}}$ to the $Q$ of FM08 (model \#1) and Wis08 (model \#2), we keep $Q$ and $Q_{\mathrm{n}}$ as two different constants for our further treatment.

With these expansions, Eq. (\ref{equ:E_tid_mod4}) involves terms in eccentricity up to order $e^8$. But since model \#4 assumes tidal locking, i.e. $\dot{E}_{\mathrm{tid}}^{\mathrm{\#4}}$ is not a function of $\Omega$, this model also yields just a lower limit for the heating rates \citep{2008Icar..193..637W}.

\subsection{Converting tidal heating into temperature increase}
\label{sub:converting}

Now that we have set up four distinct models for the calculations of the additional tidal heating term for the BDs, there are two physical processes that will be driven by these energy rates: tidal inflation and temperature increase. Let's take $\bar{L}$ as the luminosity of either of the two 2M0535$-$05 BDs that it would have if it were a single BD and $\bar{R}$ and $ \bar{T}_{\mathrm{eff}}$ as its corresponding radius and effective temperature. Then, by the Stefan-Boltzmann law \citep{Stefan, 1884AnP...258..291B}

\begin{equation}\label{equ:StefanBotzlmann}
  \bar{L} = 4 \pi \bar{R}^2 \sigma_{\mathrm{SB}} \bar{T}_{\mathrm{eff}}^4 ,
\end{equation}

\noindent
where $ \sigma_{\mathrm{SB}}$ is the Stefan-Boltzmann constant. The radial expansion in the binary case is given by $\mathrm{d}R = R - \bar{R}$ and the temperature increase by $\mathrm{d}T = T_{\mathrm{eff}} - \bar{T}_{\mathrm{eff}}$. In its present state, the BD has a luminosity

\begin{equation}\label{equ:L}
  L = \dot{E}_{\mathrm{in}} + \bar{L} ,
\end{equation}

\noindent
where $\dot{E}_{\mathrm{in}}$ is some additional internal energy rate. Solving Eq. (\ref{equ:L}) for the temperature increase yields:

\begin{equation}\label{equ:dT1}
  \mathrm{d}T = \left( \frac{\dot{E}_{\mathrm{in}}}{4 \pi R^2 \sigma_{\mathrm{SB}}} + \left[ \frac{\bar{R}}{R} \right]^2 \bar{T}_{\mathrm{eff}}^4 \right)^{1/4} - \bar{T}_{\mathrm{eff}} .
\end{equation}

\begin{figure}
\centering
  \scalebox{0.52}{\includegraphics{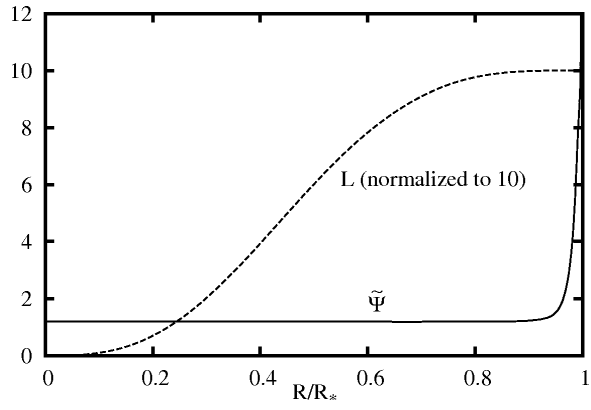}}
  \vspace{0.25cm}
  \caption{Degeneracy parameter $\tilde{\Psi} = k_{\mathrm{B}} T / (k_{\mathrm{B}} T_{\mathrm{F}})$ (solid line) with model parameters similar to those of the 2M0535$-$05 primary and radius-integrated luminosity $L$ (dashed line) as a function of radius. To fit into the plot, $L$ is normalized to 10.}
  \label{fig:psi}
\end{figure}

In the next step, we quantify the amount of tidal energy that is converted into internal energy, leading to an increase in effective temperature. Since we will use the virial theorem for an ideal, monoatomic gas to estimate the partition between internal and gravitational energy, we first have to assess the adequacy of treating the 2M0535$-$05 BDs as ideal gases. We therefore show the degeneracy parameter $\tilde{\Psi} = k_{\mathrm{B}} T / (k_{\mathrm{B}} T_{\mathrm{F}})$ as a function of radius in Fig.~\ref{fig:psi} \citep[][I. Baraffe, private communication.]{2000ARA&A..38..337C}. Here, $k_{\mathrm{B}}$ is the Boltzmann constant, $T$ is the local temperature within the gas and $E_{\mathrm{F}} = k_{\mathrm{B}} T_{\mathrm{F}}$ is the Fermi energy of a partially degenerate electron gas with an electron Fermi temperature $T_{\mathrm{F}}$. With respect to $M, T_\mathrm{eff}$ and $\log(g)$, $g$ being the body's gravitational acceleration at the surface, the BD structure model corresponds to that of the primary, but with an age of 4.9\,Myr. We find that for most of the BD, i.e. that portion of the structure in which the majority of the luminosity is released, $\tilde{\Psi}$ is of order 1. This means that we may indeed approximate the BDs as ideal gases.

With the time derivative of the virial theorem for an ideal monoatomic gas \citep[][Sect. 3.1 therein]{1990sse..book.....K},

\begin{equation}
L = \dot{E}_{\mathrm{in}} = - \dot{E}_{\mathrm{G}} / 2 ,
\end{equation}

\noindent
where $\dot{E}_{\mathrm{G}}$ is the temporal change in gravitational energy, we find that half of the additional tidal energy is converted into internal energy and the other half causes an expansion of the BD. There are currently no models for tidal inflation in BDs and the treatment is beyond the scope of this paper. Instead of including the modeled BD radii $\bar{R_i}$ into Eq. (\ref{equ:dT1}) we avoid further uncertainties and fix $\bar{R}/R = 1$ (see Sect. \ref{sec:conclusions} for a discussion of tidal inflation in the evolutionary context). The increase in effective temperature due to tidal heating then becomes

\begin{equation}\label{equ:dT2}
  \mathrm{d}T = \left( \frac{\dot{E}_{\mathrm{tid}} / 2}{4 \pi R^2 \sigma_{\mathrm{SB}}} + \bar{T}_{\mathrm{eff}}^4 \right)^{1/4} - \bar{T}_{\mathrm{eff}} .
\end{equation}

\noindent
For $\bar{T}_{\mathrm{eff},i}$ we took the values predicted by the \citet{1998A&A...337..403B} models (see Table \ref{tab:parameters}).

Our neglect of tidal inflation makes this temperature increase an upper limit. Given that this neglect is arbitrary, we estimate how our constraints for $\log(Q_2) = 3.5$ and $\psi_2 = 50^\circ$ would change, if tidal inflation played a role in 2M0535$-$05. Comparing the observed radii of both BDs with the model predictions (see Table \ref{tab:parameters}), radial expansions of 10\% for the primary and 20\% for the secondary seem realistic. Theoretical investigations of tidal heating on the inflated transiting planet HD209458b by \citep{2009ApJ...700.1921I} support an estimate of tidal inflation by 20\%. As a test, we assumed that the secondary BD in 2M0535$-$05 is tidally inflated, where its radius in an isolated scenario would be 80\% of its current value, i.e. $\bar{R} = 0.8 \cdot R$ in Eq. (\ref{equ:dT1}). In the non-inflated scenario with $\bar{R}/R = 1$, the BD would reach a temperature increase of d$T = 60$\,K at $\log(Q_2) = 3.5$ and $\psi_2 = 50^\circ$ with model \#2 (see Sect. \ref{sub:tid_mod2_results}). With the inflation, however, $\log(Q_2) \approx 2.7$ is needed to achieve the same heating at $\psi_2 = 50^\circ$, whereas no obliquity at $\log(Q_2) = 3.5$ would yield significant heating. Thus, if tidal inflation in the secondary BD increases its radius by 20\%, then the value for the dissipation function required to yield the same $T_\mathrm{eff}$ would be about 0.8 smaller in $\log(Q)$ than in the case of no inflation. Therefore, the temperature we report in Sect. \ref{sec:results} may, at worst, correspond to $\log(Q)$ that is smaller by 0.8.

\section{Results}
\label{sec:results}

\subsection{Orbital evolution}
\label{sub:orb_evol}

In order to get a rough impression of how far the orbital configuration of the system has evolved, we used the equations given in FM08, to compute the change of its eccentricity $e$ and of a possible obliquity $\psi_2$ of the secondary within the last 1.5\,Myr. Since this time span is the upper bound for the system's age, confined by its localization within the Orion Nebulae and indicated by comparison with BD evolutionary tracks, we thus get the strongest changes in $e$ and $\psi_2$. If any initial obliquity would be washed out already, $\psi_i$ could be neglected in the calculations of tidal heating. Furthermore, the measured eccentricity $e$ could give a constraint to the tidal dissipation function $Q$. Computations based on the theory of `constant-time-lag' yield qualitatively similar results.

For the evolution of $e$, we relied on Eq. (\ref{equ:e_FM}). We took the observed eccentricity $e = 0.3216$ as a starting value and evolved it backwards in time. To evolve the system into the past, we changed the sign of the right side of the equation. Furthermore, we assumed that the quality factors $Q_1$ and $Q_2$ of the primary and secondary are equal, leading to $Q_1 = Q_2  \eqqcolon \tilde{Q}$ and  $\tilde{\varepsilon}_{k,i} \coloneqq \tilde{\varepsilon}_{k,i}(\Omega_i,n,\tilde{Q})$, because we are merely interested in a tentative estimate so far. This assumption should be a good approximation due to the similarity of the both components in terms of composition, temperature, mass, and radius.

\begin{figure*}
  \centering
  \scalebox{0.52}{\includegraphics{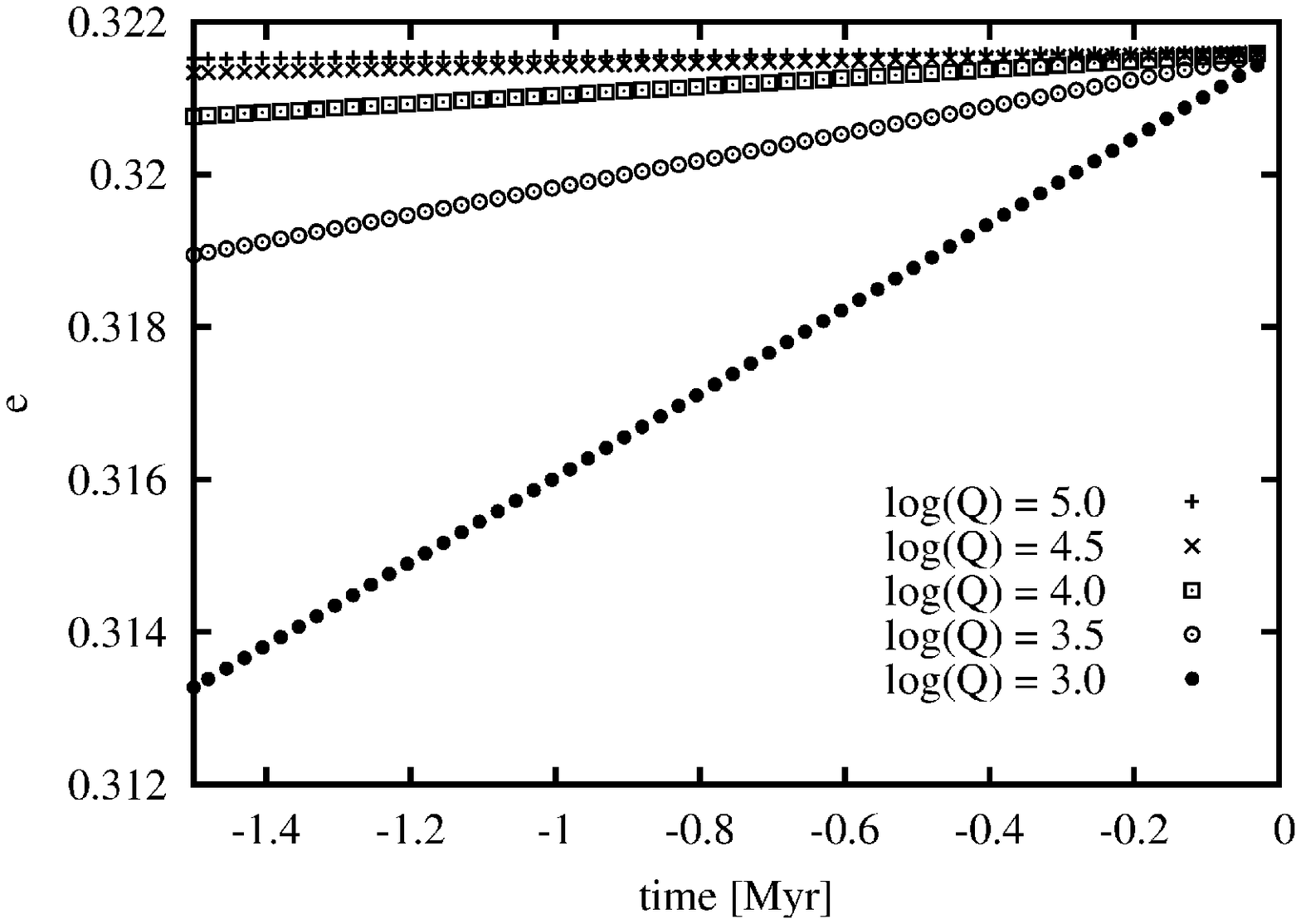}}
  \hspace{0.8cm}
  \scalebox{0.52}{\includegraphics{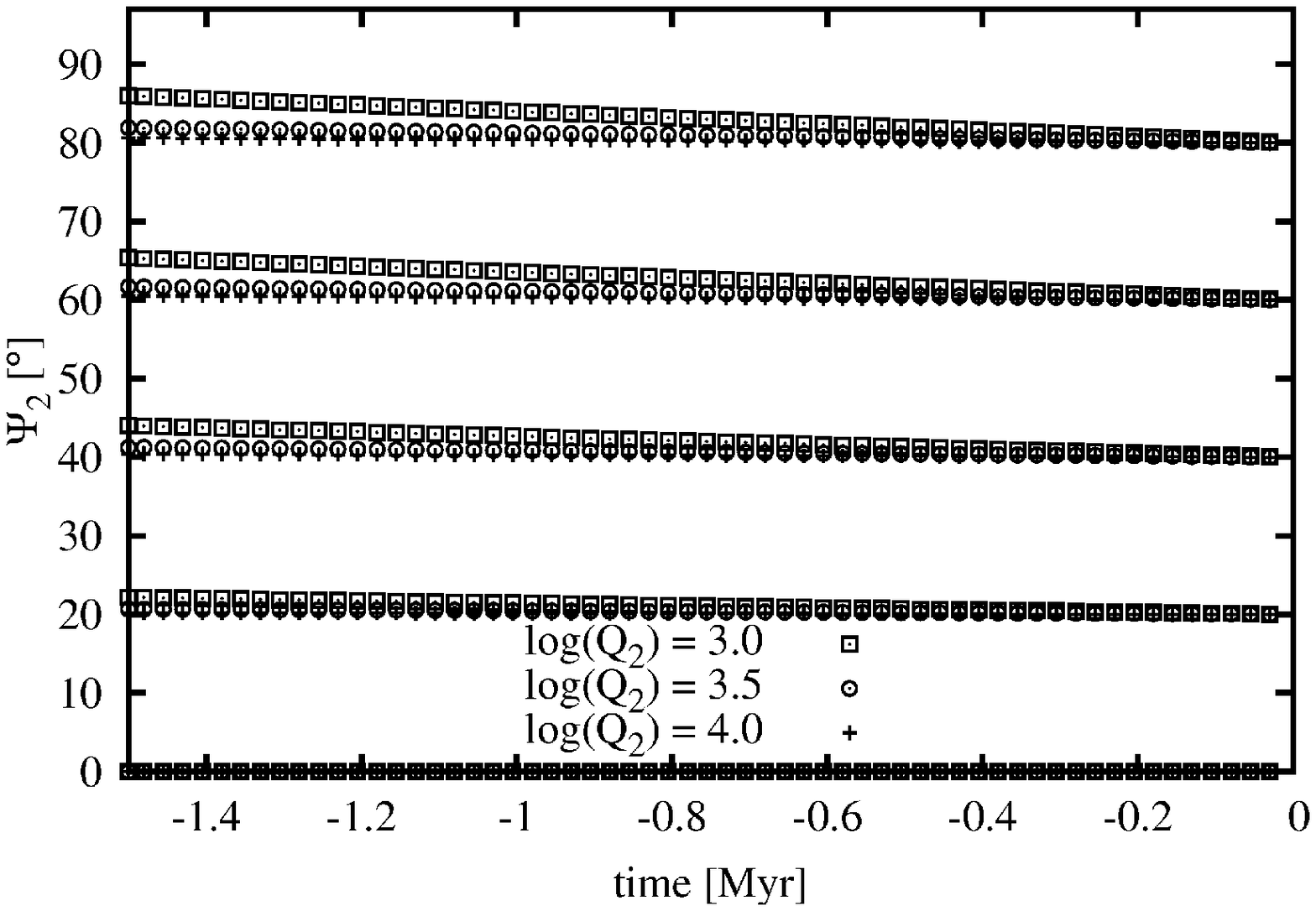}}
  \vspace{0.25cm}
  \caption{Orbital evolution of 2M0535$-$05 after model \#1 going back in time for 1.5\,Myr. \textit{Left:} Eccentricity evolution. Depending on $\tilde{Q}$ and on the age of the system, its initial eccentricity has not been smaller than $\approx 0.3133$, which is $\approx 97.4\%$ of its current value. \textit{Right}: Obliquity evolution of the secondary BD for three different values of $Q_2$. Simulations started at `time = 0' for $\psi_2 \in \{0^\circ, 20^\circ, 40^\circ, 60^\circ, 80^\circ \}$ and were evolved backwards in time. For $\log(Q_2) > 4$ there is no significant change in $\psi_2$. For all the treated values of $Q_2$, the obliquity of the 2M0535$-$05 secondary is still close to its natal state.}
 \label{fig:ecc_I2}
\end{figure*}

The observed eccentricity of the system might give a constraint to the possible values for $\tilde{Q}$ since $\mathrm{d}e/\mathrm{d}t$ depends on $\tilde{Q}$ via $\tilde{\varepsilon}_{k,i}$. Certain $\tilde{Q}$ regimes could be incompatible with the observed eccentricity of the system at a maximum age of 1.5\,Myr, if these $\tilde{Q}$ values would have caused the eccentricity to decay rapidly to 0 within this time. However, our simulations (Fig.~\ref{fig:ecc_I2}) show that the system has not yet evolved very far for the whole range of $\tilde{Q}$ and that the eccentricity of 2M0535$-$05 is in fact increasing nowadays. In this system, circularization does not occur. The observed eccentricity of 0.3216 consequently does not constrain $\tilde{Q}$. In this first estimate, we fixed all other parameters in time, i.e. we neglected an evolution of the semi-major axis $a$, of possible obliquities $\psi_{i}$ and we used constant radii $R_{i}$ and rotational frequencies $\Omega_{i}$. We did this because we cannot yet incorporate the evolutionary behavior of the components' radii $R_{i}$ in the context of tides and furthermore, there is no knowledge about possible misalignments $\psi_{i}$ between the orbital plane and the equatorial planes of the primary and secondary, respectively. A consistent evolution of $R_i$, however, is necessary to evolve $\mathrm{d}a/\mathrm{d}t$ as a function of $\psi_1$ and $\psi_2$, as given by Eq. (\ref{equ:a_FM}). Such a calculation was beyond the scope of this study.

The relative spin-geometry of the two BD rotational axes with respect to the orbital plane and with respect to each other is unknown in 2M0535$-$05. Anyhow, we can estimate if a possible obliquity that once existed for one of the BDs would still exist at an age of 1.5\,Myr or if it would have been washed out up to the present. We used a numerical integration of Eq. (\ref{equ:psi_FM}) to evolve $\psi_2$ forward in time (Fig.~\ref{fig:ecc_I2}). For the secondary's initial obliquity $\psi_{\mathrm{ini},2}$, we plot the state of $\psi_2$ as a function of the quality factor $Q_2$ after an evolution of 1.5\,Myr. We see that even for a very small quality factor of $10^3$ and high initial obliquities the secondary is basically in its natal configuration today. Thus, it is reasonable to include a putative misalignment of the secondary with respect to the orbital plane in our considerations. As shown below, this is crucial for the calculations of the tidal heating and the temperature reversal.

\subsection{Tidal heating in 2M0535$-$05 with model \#1}
\label{sub:tid_mod1_results}

In Fig. \ref{fig:E_mod1}, we show the results for the tidal heating rates as computed after tidal model \#1. As given by Eq. (\ref{equ:E_tid_mod1_2M}), the tidal heating of the primary does not depend on a putative obliquity, whereas that of the secondary does. Using this model, we find that the luminosity gain of the secondary is, over the whole $Q$ range, smaller than that of the primary, which mainly results from the relation $\dot{E}_{\mathrm{tid},i}^{\mathrm{\#1}} \propto R_i^5$. Figure \ref{fig:E_mod1} also shows that a growing obliquity shifts the gain in thermal energy towards higher values for a fixed $Q_2$. The observed overshoot of $\approx 10^{24}$\,W in the secondary's luminosity can be reproduced with very small quality factors of $Q_2 \approx 10^3$ and high obliquities up to $\psi_2 \approx 90^{\circ}$.

In Fig. \ref{fig:dT_mod1}, we show the results for the temperature increase as per Eq. (\ref{equ:dT2}) with the tidal energy rates coming from model \#1. These rates yield only a slight temperature increase for both constituents. Even for low $Q$ values of order $10^{4}$ and high obliquities of the secondary, the heating only reaches values $\lesssim 10$\,K. We also see that the heating for the primary is computed to be greater than that for the secondary and no temperature reversal would be expected. If both BDs have the same Q values, then model \#1 is unable to explain the temperature reversal. We cannot rule out a system in which, e.g., $Q_1 = 10^5$ and $Q_2 = 10^3$, for which model \#1 could explain the reversal. However, there is no reason to expect that similar bodies have $Q$ values that span orders of magnitude. Hence, we conclude that model \#1 can neither reproduce the luminosity overshoot of the secondary nor the system's temperature reversal.

\subsection{Tidal heating in 2M0535$-$05 with model \#2}
\label{sub:tid_mod2_results}

This model yields the highest heating rates and hence temperature increases. The contrast between the absolute energy rates within the primary $\dot{E}_{\mathrm{tid,1}}^{\mathrm{\#2}}$ and the secondary $\dot{E}_{\mathrm{tid,2}}^{\mathrm{\#2}}$ is very small. In fact, for any given point in $\psi$-$Q$ space, the heating rates differ only by $\log(\dot{E}_{\mathrm{tid,1}}^{\mathrm{\#2}}/\mathrm{W})-\log(\dot{E}_{\mathrm{tid,2}}^{\mathrm{\#2}}/\mathrm{W}) \approx 0.1$ (Fig. \ref{fig:E_mod2}). The tidal energy rates of the secondary become comparable to the observed luminosity overshoot at $\log(Q_2) \approx 3.5$ and $\psi_2 \approx 50^\circ$, where $\dot{E}_{\mathrm{tid,2}}^{\mathrm{\#2}} \approx 10^{24}$\,W. A comparison of the heating rates from model \#2 with those of model \#1 for either of the BDs shows that model \#2 provides higher rates, with growing contrast for increasing obliquities.

The temperature increase arising from the comparable heating rates is inverted for a given spot on the $\psi$-$\log(Q)$ plane. If both BDs had the same obliquity and the same dissipation factor, the secondary would experience a higher temperature increase. As presented in Fig. \ref{fig:dT_mod2}, the temperature increase after model \#2 is significant only in the regime of very low $Q$ and high obliquities. Neglecting any orbital or thermal evolution of the system, the observed temperature reversal could be reproduced by assuming an obliquity for the secondary while the primary's rotation axis is nearly aligned with the normal of the orbital plane. We note that the real heating will probably be greater since model \#2 assumes synchronous rotation, which is not the case for both BDs in 2M0535$-$05 (see Table \ref{tab:parameters}). The values of $Q_2$ and $\psi_2$ necessary to account for the observed increase in $L_2$ and $T_{\mathrm{eff},2}$ may thus be further shifted towards more reasonable numbers, i.e. $Q_2$ might also be higher than $10^{3.5}$ and the obliquity might be smaller than $50^\circ$. Thus, for a narrow region in the $\psi$-$\log(Q)$ plane, model \#2 yields tidal energy rates for the secondary comparable to its observed luminosity overshoot and in this region the computed temperature increase can explain the observed temperature reversal.

\subsection{Tidal heating in 2M0535$-$05 with model \#3}
\label{sub:tid_model_3_results}

Since the only free parameter in this model is the putative fixed time lag $\tau$, we show the tidal heating rates for both the primary and the secondary only as a function of $\tau$ in Fig. \ref{fig:E_mod3} with $0\,\mathrm{s} < \tau < 300\,\mathrm{s}$. For this range, model \#3 yields energy rates and temperature rises that are compatible with the observed luminosity and temperature overshoot of the secondary. For $\tau \gtrsim 100$\,s the heating rate for the secondary becomes comparable to the observed one, namely $\dot{E}_{\mathrm{tid},2}^{\mathrm{Hut}} \approx 10^{24}$\,W. However, assuming a similar time lag $\tau_1$ for the primary, the luminosity gain of the primary BD would be significantly higher than that of the secondary, which is not compatible with the observations. The assumption of $\tau_1 \approx \tau_2$ should be valid since both BDs are very similar in their structural properties, such as mass, composition, temperature, and radius.

The corresponding temperature increase is plotted in Fig. \ref{fig:dT_mod3}. It shows that the more massive BD would experience a higher temperature increase than its companion, assuming similar time lags. Since tidal heating is underway in 2M0535$-$05 and was probably similar in the past (see Sect. \ref{subsub:tid_model_1}), tidal heating after model \#3 would have been more important on the primary, forcing it to be even hotter than it would be without the perturbations of the secondary. The temperature difference between the primary and the secondary, which is anticipated by BD evolutionary models, would be even larger. Thus, the temperature inversion cannot be explained by tidal model \#3.

\subsection{Tidal heating in 2M0535$-$05 with model \#4}
\label{sub:tid_model_4_results}

The calculations based on model \#4 yield significant heating rates in both BDs. Like in the case of models \#1 and \#2, the luminosity gain of the secondary at a fixed obliquity is, over the whole $Q_{\mathrm{n}}$ range, smaller than that of the primary (Fig. \ref{fig:E_mod4}). As for model \#2, the difference between $\dot{E}_{\mathrm{tid},1}^\mathrm{\#4}(\psi)$ and $\dot{E}_{\mathrm{tid},2}^\mathrm{\#4}(\psi)$ is less pronounced than in model \#1. Assuming spin-orbit alignment for the primary and a pronounced obliquity of the secondary, tidal heating rates of $\dot{E}_{\mathrm{tid},2}^{\mathrm{\#4}} = 10^{24}$\,W can be reached with $\log(Q_{\mathrm{n},2}) \approx 3.5$ and $\psi_2 \approx 50^\circ$. 

Like model \#2, \#4 produces a reversal in temperature increase by means of the modified Stefan-Boltzmann relation in Eq. (\ref{equ:dT2}), due to the comparable heating rates of both BDs and the significantly smaller radius of the secondary (Fig. \ref{fig:dT_mod4}). We find a reversal in tidal heating, i.e. $\mathrm{d}T_2 > \mathrm{d}T_1$ for any given point in $\psi$-$Q_{\mathrm{n}}$ space. A temperature increase of $\gtrsim 40$\,K can be reached with $\log(Q_{\mathrm{n},2}) \approx 3.5$ and $\psi_2 \approx 50^\circ$. Since the equations of model \#4 provide merely a lower limit due to the assumption of asymptotic non-synchronous rotation, $Q_{\mathrm{n},2}$ might also be higher than $10^{3.5}$ and the obliquity might be smaller than $50^\circ$. Similar to model \#2, tidal model \#4 can reproduce the observed temperature reversal in a narrow region of the $\psi$-$\log(Q)$ parameter space.

\section{Discussion}
\label{sec:discussion}

We employed several tidal models to explore the tidal heating in 2M0535$-$05. We found that, assuming similar tidal quality factors $Q$ and obliquities $\psi$ for both BDs, the constant-phase-lag model \#2 and the constant-time-lag model \#4 yield a stronger increase in effective temperature on the secondary mass BD than on the primary. For certain regimes of $Q_2$ and $\psi$, the tidal energy rates in the secondary are of the correct amount to explain the larger temperature in the smaller BD. A comparison between our computations based on the models \#1 and \#2 on the one hand and \#3 and \#4 on the other hand is difficult. The reference to a fixed tidal time lag might only be reconciled with the assumption of $Q_{\mathrm{n}}^{-1} = n \ \tau$ as done by Lev07, which is at least questionable since the assumption of a fixed time lag is not compatible with a fixed phase lag. Furthermore, model \#3 does not invoke obliquities, which also complicates direct comparisons of the model output.

\subsection{Constraints on the tidal dissipation function for BDs, $Q_{\mathrm{BD}}$}
\label{sec:constraints_Q}

\subsubsection{The Rossiter-McLaughlin effect in 2M0535$-$05}
\label{subsub:rme}

The geometric implication of the most promising tidal models \#2 and \#4 is that the obliquity of the 2M0535$-$05 primary is negligible and that of the secondary is $\psi_2 \approx 50^\circ$ -- provided tidal heating accounts for the $T_{\mathrm{eff}}$ reversal and the luminosity excess of the secondary. There does exist an observational method to measure the geometric configuration of eclipsing systems, called the Rossiter-McLaughlin effect (RME) \citep{1924ApJ....60...15R, 1924ApJ....60...22M}.

The RME appears during transits in front of rotating stars. Hiding a fraction of the star's surface results in the absence of some corresponding rotational velocity contribution to the broadening of the stellar lines. Thus, the changes in the line profiles become asymmetric (except for the midpoint of the transit) and the center of a certain stellar line is shifted during a transit, which induces a change of the star's radial velocity. The shape of the resulting radial velocity curve depends on the effective area covered by the transiting object and its projected path over the stellar surface with respect to the spin axis of the covered object \citep[for a detailed analysis of the RME see][]{2005ApJ...622.1118O}.

\begin{figure*}[h]
  \centering
  \scalebox{0.49}{\includegraphics{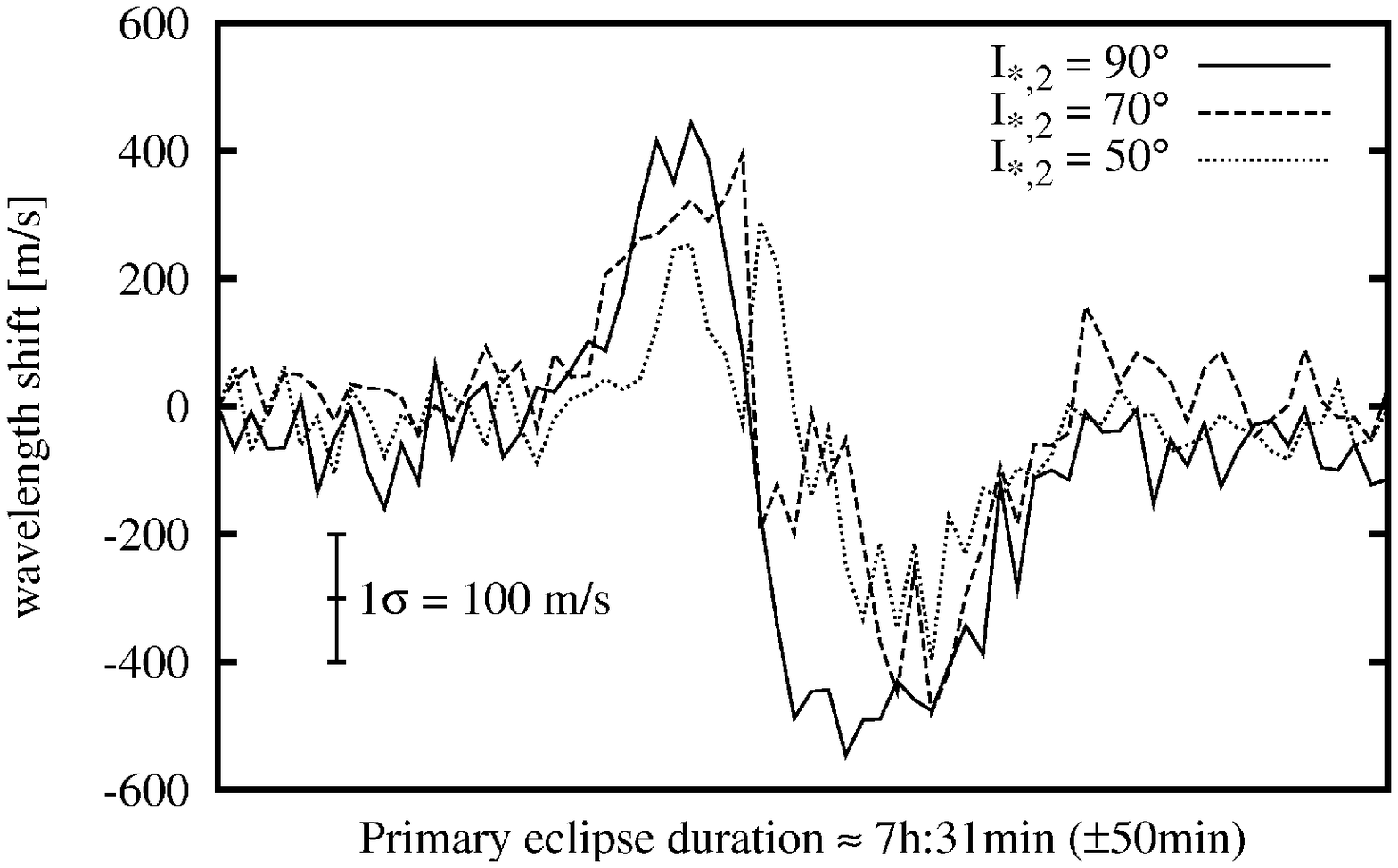}}
  \hspace{0.8cm}
  \scalebox{0.49}{\includegraphics{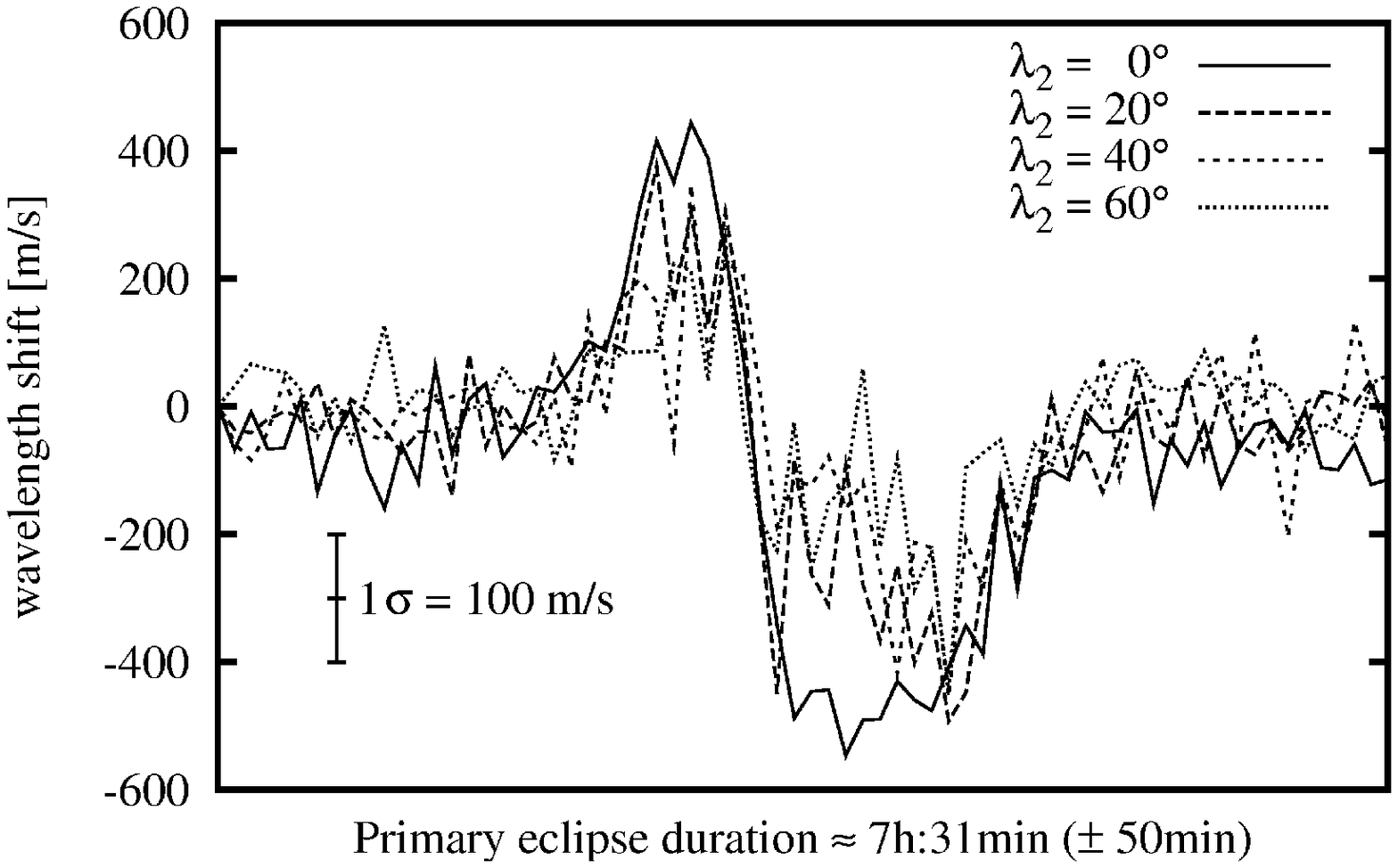}}
  \vspace{0.25cm}
  \caption{Simulations for the Rossiter-McLaughlin effect as it would be seen with UVES during the primary eclipse of 2M0535$-$05, which occurs when the secondary mass BD is occulted by the primary. The S/N is 7. \textit{Left}: The orbital inclination $i$ is fixed at $88.49^\circ$ (see Table \ref{tab:parameters}) and $\lambda = 0$, which means the transiting primary BD follows a path parallel to the secondary's equator. The alignment of the secondary's spin axis $I_{\star,2}$ varies between $90^\circ$ (perpendicular to the line of sight) and $50^\circ$. \textit{Right}: With $i$ fixed at $88.49^\circ$ and $I_{\star,2} = 90^\circ$, $\lambda_2$ varies between $0^\circ$ (primary path parallel to the secondary's equator) and $60^\circ$ (primary path strongly misaligned with the secondary's equator).}
 \label{fig:RM}
\end{figure*}

Using a code originally presented in \citet{2009A&A...499..615D}, we have undertaken simulations of the RME for various geometric configurations of 2M0535$-$05 during the primary eclipse\footnote{The `primary eclipse' refers to the major flux decrease in the system's light curve. Due to the significantly higher effective temperature of the secondary mass BD the primary eclipse occurs when the primary mass component transits in front of the secondary companion, as seen from Earth.} as it would be seen with the Ultraviolet and Visual Echelle Spectrograph (UVES) at the Very Large Telescope (VLT) (see Fig.~\ref{fig:RM}). For the data quality we assumed the constraints given by the UVES at the VLT exposure time calculator\footnote{http://www.eso.org/observing/etc} in version 3.2.2. The computations show that, using Th-Ar reference spectra and also the telluric A and B bands as benchmarks, a time sampling with one spectrum every 1\,245\,s and a S/N of $\gtrsim$ 7 around 8\,600\,$\AA$ are necessary to get 21 measurements during the primary eclipse and an accuracy of $\lesssim$ 100\,m/s.

In principle, there are four parameters for the background object of the transit to be fitted in our simulations of the RME: the rotational velocity $v_{\mathrm{rot}}$, the inclination of the spin axes with respect to the line of sight $I_{\star}$, the angle between the projection of the spin and the projection of the orbital plane normal onto the celestial plane $\lambda$, and the orbital inclination with respect to the line of sight $i$. From light curve analyses, both rotational velocities in 2M0535$-$05 and the orbital inclination $i$ are known. Thus, for the simulation of the primary eclipse $I_{\star,2}$ and $\lambda_2$ are the remaining free parameters.

The obliquities $\psi_{i \ | \ i = 1, 2}$, i.e. the real 3-dimensional angle between the orbital normal and the spin axis of the occulted object, is related to the other angles as

\begin{equation}\label{equ:psi_definition}
  \cos(\psi_i) = \cos(I_{\star,i}) \cos(i) + \sin(I_{\star,i}) \sin(i) \cos(\lambda_i) .
\end{equation}

\noindent
While the two obliquities $\psi_i$ are intrinsic angles of the system, they cannot be measured directly. They can only be inferred from $i$, $I_{\star,i}$ and $\lambda_i$, which depend on the position of the observer with respect to the system. Since we are only interested in the possible options for the measurement of the obliquities in 2M0535$-$05, we refer the reader to the paper by \citet{2005ApJ...631.1215W} for a discussion of Eq. (\ref{equ:psi_definition}) and the geometrical aspects of the RME. With $i = 88.49^{\circ}$ the first term in Eq. (\ref{equ:psi_definition}) degrades to insignificance, which yields $\cos(\psi_i) \approx \sin(I_{\star,i}) \cos(\lambda_i)$.

At low values for $I_{\star,i}$ and $\lambda_i$ the fitted solutions to the RME are degenerate and there are multiple solutions within a certain confidence interval. But our simulations for the transit show that the error due to the observational noise is on the same order as the error due to degeneracy and thus we find standard deviations in $I_{\star,2}$ and $\lambda_2$ of $\sigma_{I_{\star,2}} \approx 20^{\circ}$ and $\sigma_{\lambda_{2}} \approx 20^{\circ}$, respectively. The uncertainty in $\psi_2$ depends not only on the uncertainties in $I_{\star,2}$ and $\lambda_{2}$ but also on the actual values of $I_{\star,2}$ and $\lambda_{2}$. But in all cases, the standard deviation in the secondary's obliquity $\sigma_{\psi_2}~<~20^\circ$.

If present in 2M0535$-$05, a considerable misalignment of the secondary BD of $50^\circ$ could be detected with a 1-$\sigma$ accuracy of $20^\circ$ or less. Thus, an observed $\psi_2$ value of $50^\circ$ would be a 2.5-$\sigma$ detection of spin-orbit misalignment. Unless RME measurements suggest $\psi \approx 90^\circ$, RME observations alone are unlikely to provide definitive evidence that any of the tidal models we consider is responsible for the temperature reversal.

\subsubsection{Further observations of BD binaries}
\label{subsub:further}

\begin{figure*}
  \centering
  \scalebox{0.52}{\includegraphics{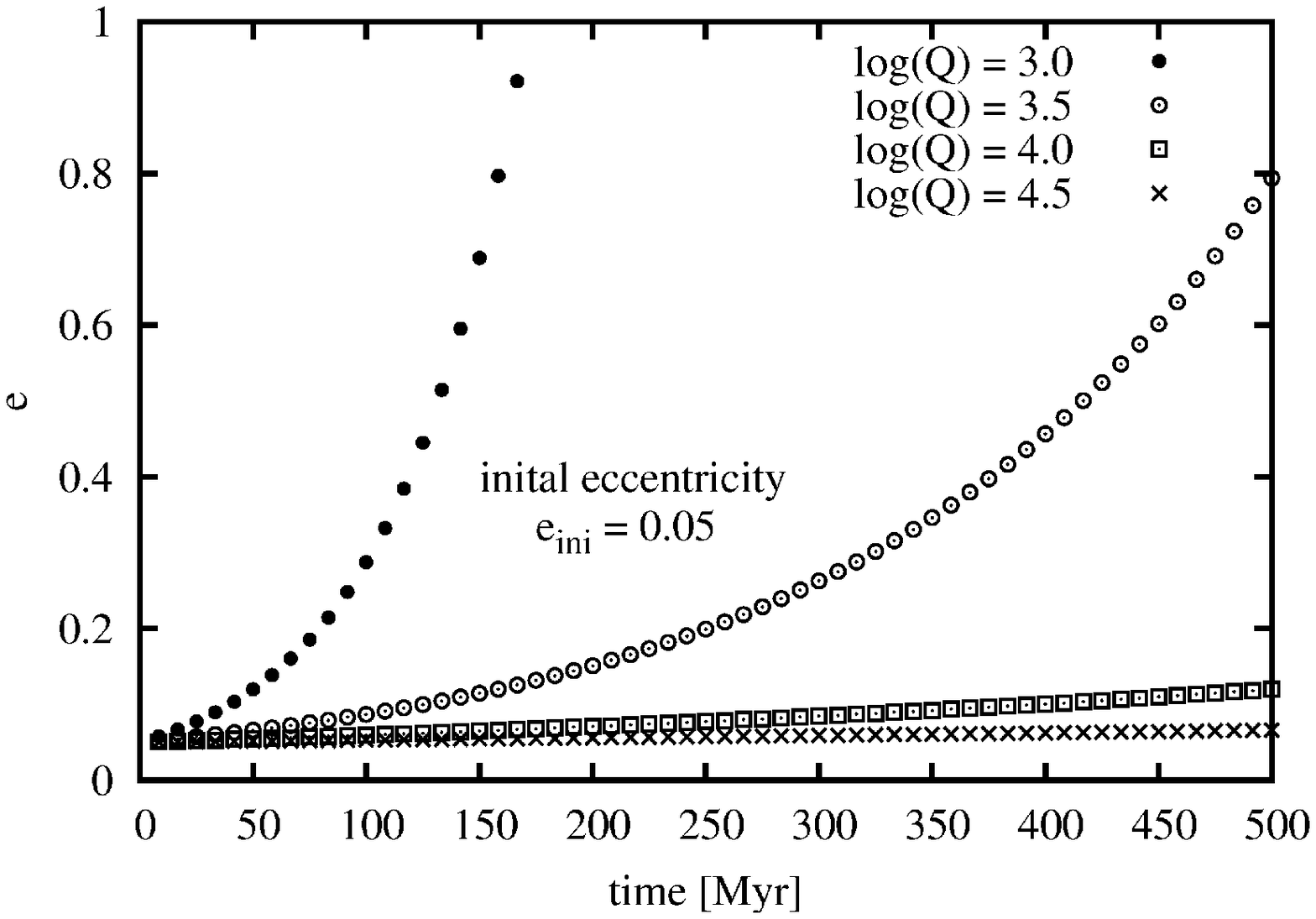}}
  \hspace{0.8cm}
  \scalebox{0.52}{\includegraphics{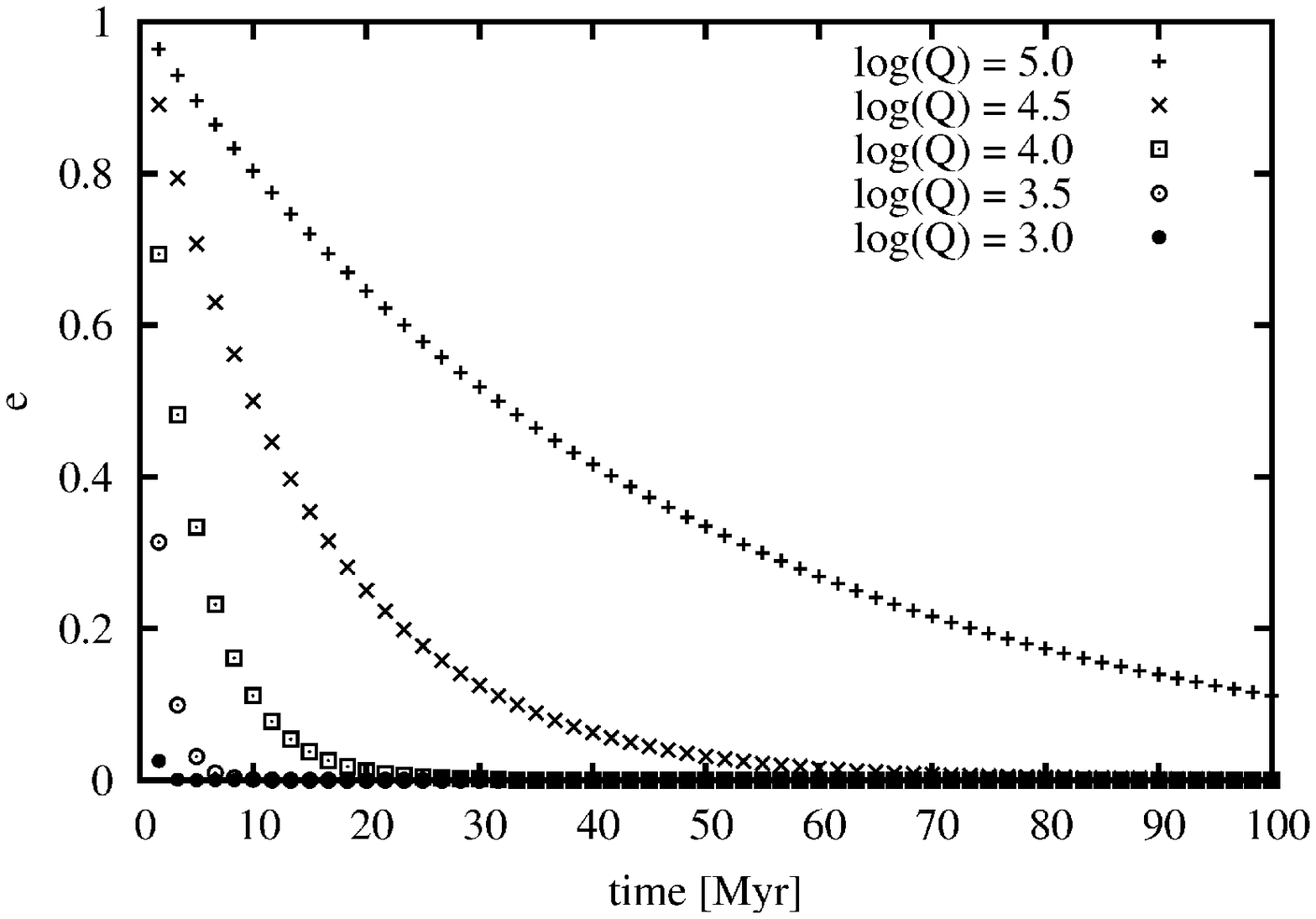}}
  \vspace{0.25cm}
  \caption{Orbital evolution of a 2M0535$-$05 analog after model \#1. \textit{Left:} Eccentricity evolution for different values of $\tilde{Q}$ for the next 500\,Myr. The initial eccentricity was arbitrarily chosen: $e_\mathrm{ini} = 0.05$. For $\log(\tilde{Q}) \lesssim 3.5$ this binary will be disrupted within 500\,Myr. \textit{Right}: Eccentricity evolution of a 2M0535$-$05 analog but with $P_1 = P_2 = 14.05$\,d for different values of $\tilde{Q}$. Contrary to the scenario in the left figure, the changed rotational period of the primary BD now leads to circularization of the system. Measurements of $e$ in LMS binaries with known ages can give lower limits to $\tilde{Q}$.}
 \label{fig:ecc_500and100Myr}
\end{figure*}

Besides the option of RME measurements for testing the geometric implications, there does exist a possibility to verify our estimate of $\log(Q) \approx 3.5$ for BDs in general. Comparison of observed orbital properties with values constrained by the equations that govern the orbital evolution might constrain the free parameters, here $Q$. Using Eq. (\ref{equ:e_FM}), we find that, assuming only a slight initial eccentricity of 0.05, the eccentricity of a BD binary system similar to 2M0535$-$05, in terms of masses, radii, rotational frequencies, and semi-major axis would increase to 1 after $\approx 500$\,Myr if the quality factors of the two BDs are $\lesssim 10^{3.5}$ (see left panel in Fig.~\ref{fig:ecc_500and100Myr}). A measurement of $e$ in such an evolved state could not constrain $Q$ in a 2M0535$-$05 analog since either the initial eccentricity could have been relatively large while the orbit evolved rather slowly due to high $Q$ values or a small initial value of $e$ could have developed to a large eccentricity due to small values of $Q$.

We also simulate the evolution of a 2M0535$-$05 analog but with a different rotational frequency of the primary constituent in order to let the eccentricity decrease with time. We neglected the evolution of all the other physical and orbital parameters since we are merely interested in a tentative estimate. For a given candidate system the analysis would require a self-consistent coupled evolution of all the differential equations. For the arbitrary case of $P_1 = P_2 = 14.05$\,d we find that, even for the most extreme but unrealistic case of an initial eccentricity equal to 1, this fictitious binary would be circularized on a timescale of 100\,Myr for $\log(\tilde{Q}) < 5$ (see right panel in Fig.~\ref{fig:ecc_500and100Myr}). Findings of old, eccentric BD binaries with rotational and orbital frequencies that yield circularization in the respective system would set lower limits to $Q$.

\subsubsection{Rotational periods in 2M0535$-$05}
\label{subsub:rotation}

Another, and in fact a crucial, constraint on $Q$ for BDs comes from the synchronization time scale $t_\mathrm{synch}$ of the two BDs in 2M0535$-$05. Following the equation given in Lev07 and taking the initial orbital mean motion and semi-major axis of the system as calculated with an uncoupled system of differential equations from model \#1, we derive $t_{\mathrm{synch},1} = 0.07$\,Myr for the primary and $t_{\mathrm{synch},2} = 0.04$\,Myr for the secondary with $\log(Q) = 3.5$. Since the rotation in both BDs is not yet synchronized with the orbit and the age of the system is about 1\,Myr, $\log(Q) = 3.5$ is not consistent with the age of 2M0535$-$05. Both components should have synchronous rotation rates already. We find the consistent value for $Q$ to be $\gtrsim 10^{4.5}$, yielding synchronization time scales $t_{\mathrm{synch},1} \gtrsim 0.69$\,Myr and $t_{\mathrm{synch},2} \gtrsim 0.37$\,Myr.

\begin{figure*}
  \centering
  \scalebox{0.52}{\includegraphics{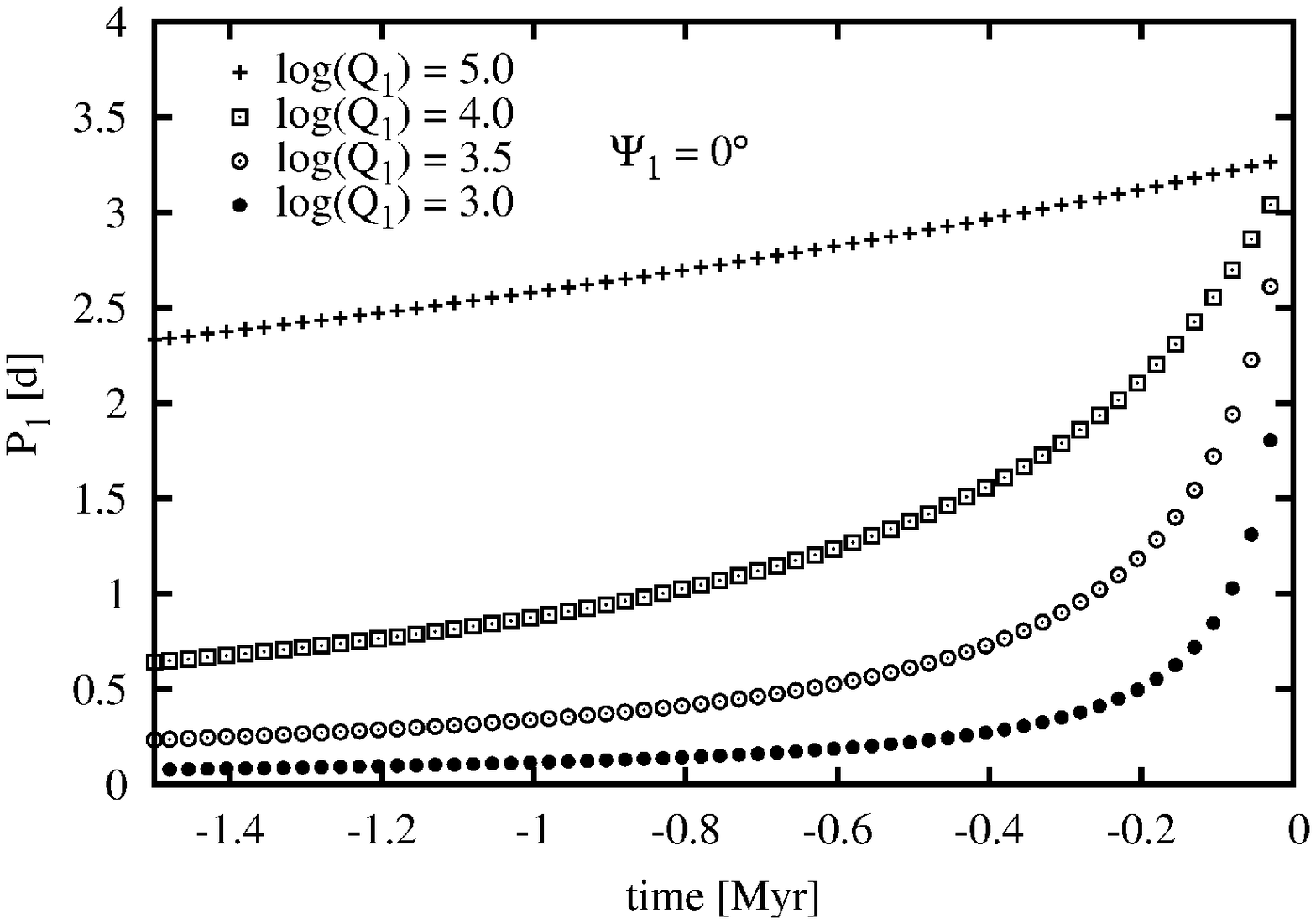}}
  \hspace{0.8cm}
  \scalebox{0.52}{\includegraphics{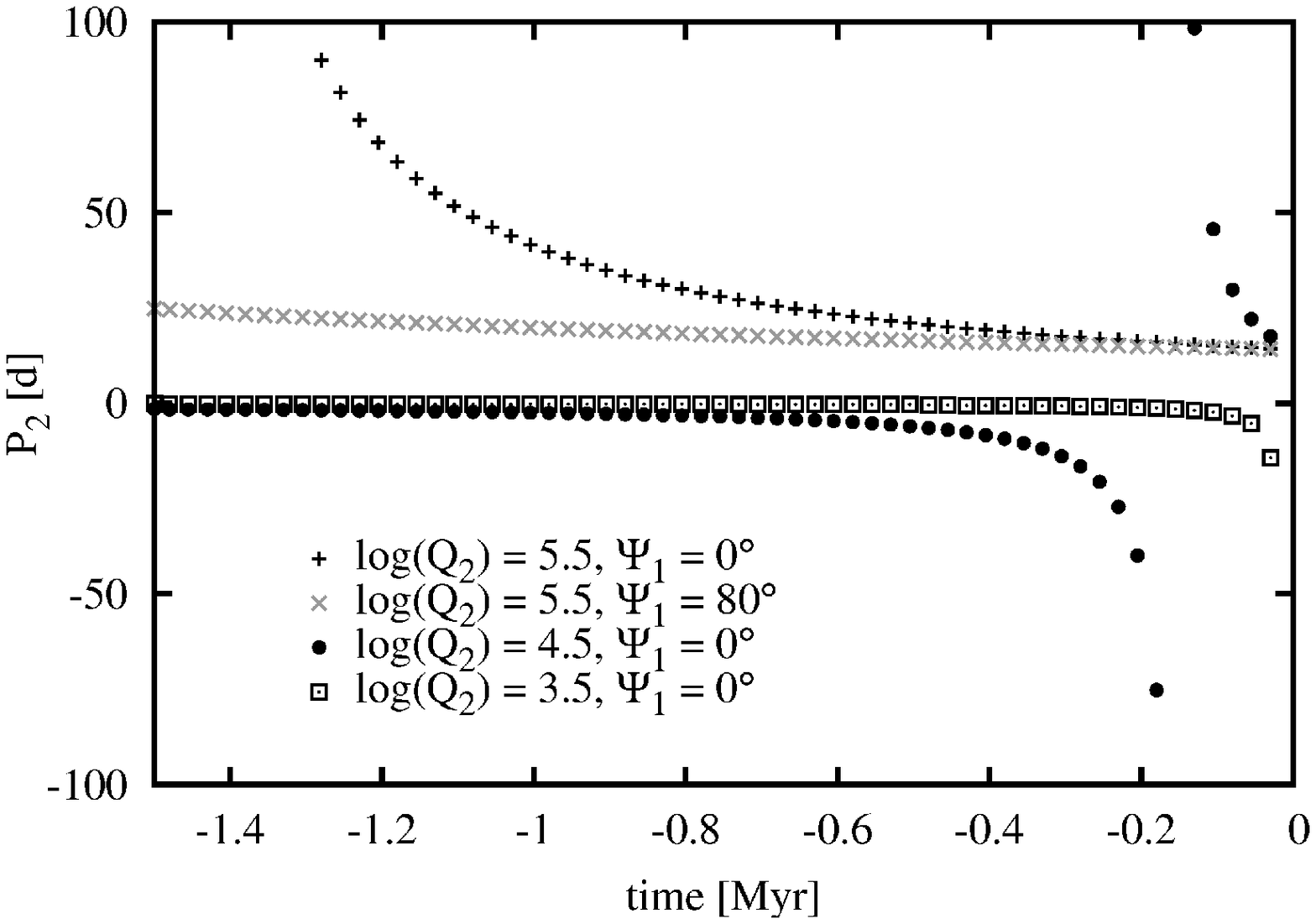}}
  \vspace{0.25cm}
  \caption{Rotational evolution of the two BDs in 2M0535$-$05 after model \#1 for different values of $Q_1$ and $Q_2$. \textit{Left:} (Primary) Going backwards in time, the rotation period decreases. For $\log(Q_1) = 3.5$, $P_1$ drops below the critical period for structural breakup of $\approx 0.5$\,d already before the date of birth around 1\,Myr ago. \textit{Right}: (Secondary) For $\log(Q_2) = 5.5$ we show the tracks for $\psi_2 = 0^\circ$ and $80^\circ$ for comparison. For $\log(Q_2) = 4.5$ the rotation direction switches at about $-0.18$\,Myr and for $\log(Q_2) = 3.5$ at roughly $-10,000$\,yr.}
 \label{fig:rot_evol}
\end{figure*}

To make this estimate for $Q$ more robust, we present the evolution of the BDs' rotational periods in Fig.~\ref{fig:rot_evol} and compare it to the critical period for a structural breakup $P_\mathrm{crit}$. The evolutionary tracks are calculated with model \#1 and Eq. (30) in FM08. As a rough approach we do not couple this equation with those for the other orbital parameters. The left panel of Fig.~\ref{fig:rot_evol} shows that for $\log(Q_1) = 3.5$ and $\psi_1 = 0^\circ$ the primary's initial rotation period 1\,Myr ago is $\approx 0.3$\,d. The initial rotation period for the secondary, for $\log(Q_2) = 3.5$ and $\psi_2 = 0^\circ$, is about -0.2\,d, where the algebraic sign contributes for a retrograde revolution (right panel in Fig.~\ref{fig:rot_evol}). For most of its lifetime, the secondary would have had a retrograde rotation and just switched the rotation direction within the last few 10,000 yr, which is very unlikely in statistical terms. Since the orbital momentum is on the order of $10^{43}$\,kg\,m$^2$\,/\,s and the individual angular momenta are about $10^{41}$\,kg\,m$^2$\,/\,s, the shrinking process might not have had a serious impact on the rotational evolution. Tides have dominated the spin evolutions.

Following \citet{2005A&A...429.1007S}, the critical breakup period $P_\mathrm{crit}$ depends only on the body's radius and its mass. The radius evolution for BDs is very uncertain for the first Myr after formation but we estimate their initial radii to be as large as the solar radius. This yields $P_{\mathrm{crit},1} \approx 0.5$\,d for both the primary and the secondary BD. As stated above, the moduli of the initial rotation periods of both BDs would have been smaller than 0.5\,d for $Q$ values of $\lesssim 10^{3.5}$. This inconsistency gives a lower limit to $Q_1$ and $Q_2$ since values of $\lesssim 10^{3.5}$ would need an initial rotation periods of both BDs which are smaller than their critical breakup periods. Obliquities larger than $0^\circ$ would accelerate the (backwards) evolution and yield even larger lower limits for $Q_1$ and $Q_2$.  Thus, our simulations of the rotational period evolution of both BDs require $\log(Q_{\mathrm{BD}}) \gtrsim 3.5$, whereas the tidal synchronization timescale even claims $\log(Q_{\mathrm{BD}}) \gtrsim 4.5$.

\subsection{Evolutionary embedment of tidal heating}
\label{sub:embedment}

Tidal heating must be seen in the evolutionary context of the system. On the one hand, the tidal energy rates generate a temperature increase on the Kelvin-Helmholtz time-scale, which is $\approx 2$\,Myr for the BDs in 2M0535$-$05 -- and thus on the order of the system's age, as per Eq. (\ref{equ:dT2}). On the other hand, tidal heating will affect the shrinking and cooling process of young BDs in terms of an evolutionary retardation. As models show \citep{1997MmSAI..68..807D, 1998A&A...337..403B, 2000ApJ...542..464C, 2000ARA&A..38..337C}, single BDs cool and shrink significantly during their first Myrs after formation. Adding an energy source comparable to the luminosity of the object will slow down the aging processes such that the observed temperature and luminosity overshoot at some later point is not only due to the immediate tidal heating but also due to its past evolution. Consequently, the luminosity and temperature overshoot in the secondary might not (only) be due to present-day tidal heating, but it could be a result of an evolutionary retardation process triggered by the presence of the primary as a perturber. Coupled radius-orbit evolutionary models have already given plausible explanations for the inflated radii of some extrasolar planets \citep{2003ApJ...588..509G, 2009ApJ...702.1413M, 2009ApJ...700.1921I, 2009arXiv0910.4394I, 2009arXiv0910.5928I}.

For a consistent description of the orbital and physical history of 2M0535$-$05, one would have to include the evolution of obliquities $\psi_i$, BD radii $R_i$, eccentricity $e$, semi-major axis $a$, and rotational frequencies $\Omega_i$. Note that there is a positive feedback between radial inflation and tidal heating: as tidal heating inflates the radius, the tidal heating rate can increase and -- in turn -- may cause the radius to inflate even more. In a self-consistent orbital and structural simulation of 2M0535$-$05, tidal inflation, neglected in our computations of the $T_{\mathrm{eff}}$ increase in Eq. (\ref{equ:dT2}), will result naturally from the additional heating term introduced by tides.

In conjunction with 2M0535$-$05 that means the actual heating rates necessary to explain the $T_{\mathrm{eff}}$ and luminosity excess in the secondary are lower than they would have to be if there would be no historical context. Relating to Figs. \ref{fig:dT_mod1}, \ref{fig:dT_mod2}, \ref{fig:dT_mod3}, and \ref{fig:dT_mod4}, the implied obliquity and $Q$ factor for the secondary are -- again -- shifted towards lower and higher values, respectively. Embedded in the historical context of tidal interaction in 2M0535$-$05, $\psi_2 < 50^{\circ}$ and $\log(Q_{2}) >3.5$ may also explain the temperature reversal and the luminosity excess of the secondary.

These trends, however, are contrary to that induced by tidal inflation. If tidal heating is responsible for a radial expansion of 10 and 20\% in the primary and secondary, the values of the dissipation factor necessary to explain the $T_\mathrm{eff}$ reversal would be $\approx 0.8$ smaller in $\log(Q_2)$ (see Sect. \ref{sub:converting}).

\section{Conclusions}
\label{sec:conclusions}

We surveyed four different published tidal models, but neglect any evolutionary background of the system's orbits and the components' radii to calculate the tidal heating in 2M0535$-$05. Our calculations based on models \#2 and \#4, which are most compatible with the observed properties of the system, require obliquities $\psi_1 \approx 0$, $\psi_2 \approx 50^{\circ}$ and a quality factor $\log(Q) \approx 3.5$ in order to explain the luminosity excess of the secondary. Additionally, the observed temperature reversal follows naturally since we may reproduce a reversal in temperature increase due to tides: $\mathrm{d}T_2 > \mathrm{d}T_1$. In model \#2, synchronous rotation of the perturbed body is assumed. Since this is not given in 2M0535$-$05, the actual heating rates will be even higher than those computed here. Our results for the heating rates as per model \#2 are thus lower limits, which shifts the implied obliquity of the secondary and its $Q$ factor to lower and higher values, respectively.

Considerations of the synchronization time scale for the BD duet and the individual rotational breakup periods yield constraints on $Q_{\mathrm{BD}}$ for BDs. We derive a lower limit of $\log(Q_{\mathrm{BD}}) > 4.5$. This is consistent with estimates of $Q$-values for M dwarfs, $\log(Q_{\mathrm{dM}}) \approx 5$, and the quality factors of Jupiter, $2 \cdot 10^{5} < Q_{\jupiter} < 2 \cdot 10^{6}$, and Neptune, $10^{4} \lesssim \log(Q_{\neptune}) \lesssim 10^{4.5}$ (see Sect. \ref{subsub:tid_model_1}). With $\log(Q_{\mathrm{BD}}) > 4.5$ tidal heating alone can neither explain the temperature reversal in the system nor the luminosity excess of the secondary.

An obliquity of $50^\circ$, however, would be reasonable in view of recent results from measurements of the RME in several transiting exoplanet systems\footnote{See http://www.hs.uni-hamburg.de/EN/Ins/Per/Heller for an overview.}. Currently, out of 18 planets there are 7 with significant spin-orbit misalignments $\gtrsim 30^\circ$ and some of them are even in retrograde orbits around their host stars. A substantial obliquity $\psi_2$ might cause an enhanced heating in the 2M0535$-$05 secondary, while the primary's spin could be aligned with the orbital spin, leading to negligible heating in the primary.

Despite the advantages of distance-independent radius and luminosity measurements of close, low-mass binaries, the comparison of fundamental properties of the constituents with theoretical models of isolated BDs must be taken with care. This applies also to the direct translation from the discrepancies between observed and modeled radii for a fixed metallicity into an apparent age difference as a calibration of LMS models \citep{2009IAUS..258..161S}. Tidal heating might be a crucial contribution to discrepancies between predicted and observed radii in other eclipsing low-mass binary systems \citep{2008MmSAI..79..562R}. As recently shown by \citet{2009ApJ...700.1921I}, tidal heating in extra-solar giant planets in close orbits at $a \lesssim 0.2$\,AU with modest to high eccentricities of $e \gtrsim 0.2$ can explain the increased radii of some planets, when embedded in the orbital history with its host star.

Improvement of tidal theories is necessary to estimate the relation between tides and the observed radii of LMS being usually too large as compared to models. A tidal model is needed for higher orders of arbitrary obliquities and eccentricities that also accounts for arbitrary rotation rates. As stated by \citet{2009ApJ...698L..42G}, a formal extension of the simple `lag-and-add' procedure of tidal frequencies the theory of constant phase lag is questionable. Besides the extension, conciliation among the various models is needed. The results from the models applied here should be considered preliminary but are suggestive and indicate the possible importance of tides in binary BD systems.

Several issues remain to be addressed for a more detailed assessment of tidal heating in 2M0535$-$05: \textit{i.} reconciliation and improvement of tidal theories; \textit{ii.} self-consistent simulations of the orbital and physical evolution of the system and the BDs; \textit{iii.} measurements of the system's geometric configuration; \textit{iv.} constraints on the tidal quality factors of BDs.

\begin{acknowledgements}

Our sincere thanks go to J. L. Bean for initiating this collaboration. The advice of S. Dreizler on the computations of the RM effect and inspirations from A. Reiners on 2M0535$-$05 and BDs in general have been a valuable stimulation to this study. We acknowledge the help of Y. G. M. Chew and K. G. Stassun on the parametrization of the 2M0535$-$05 BD binary and we appreciate the contribution from I. Baraffe to the modeling of the BDs' structures. The referee Jean-Paul Zahn deserves our honest gratitude for his crucial remark on the tidal synchronization time scale. R. Heller is supported by a PhD scholarship of the DFG Graduiertenkolleg 1351 ``Extrasolar Planets and their Host Stars''. R. Barnes acknowledges funding from NASA Astrobiology Institute's Virtual Planetary Laboratory lead team, supported by NASA under Cooperative Agreement No. NNH05ZDA001C. R. Greenberg, B. Jackson, and R. Barnes were also supported by a grant from NASA's Planetary Geology and Geophysics program. This research has made use of NASA's Astrophysics Data System Bibliographic Services. 

\end{acknowledgements}

\bibliographystyle{aa} 
\bibliography{2009-1_Heller_BD_2MASS_TIDES}

\begin{thebibliography}{58}
\expandafter\ifx\csname natexlab\endcsname\relax\def\natexlab#1{#1}\fi

\bibitem[{{Aksnes} \& {Franklin}(2001)}]{2001AJ....122.2734A}
{Aksnes}, K. \& {Franklin}, F.~A. 2001, \aj, 122, 2734

\bibitem[{{Baraffe} {et~al.}(1998){Baraffe}, {Chabrier}, {Allard}, \&
  {Hauschildt}}]{1998A&A...337..403B}
{Baraffe}, I., {Chabrier}, G., {Allard}, F., \& {Hauschildt}, P.~H. 1998, \aap,
  337, 403

\bibitem[{{Baraffe} {et~al.}(2002){Baraffe}, {Chabrier}, {Allard}, \&
  {Hauschildt}}]{2002A&A...382..563B}
{Baraffe}, I., {Chabrier}, G., {Allard}, F., \& {Hauschildt}, P.~H. 2002, \aap,
  382, 563

\bibitem[{{Barnes} {et~al.}(2009){Barnes}, {Jackson}, {Raymond}, {West}, \&
  {Greenberg}}]{2009ApJ...695.1006B}
{Barnes}, R., {Jackson}, B., {Raymond}, S.~N., {West}, A.~A., \& {Greenberg},
  R. 2009, \apj, 695, 1006

\bibitem[{{Bodenheimer} {et~al.}(2001){Bodenheimer}, {Lin}, \&
  {Mardling}}]{2001ApJ...548..466B}
{Bodenheimer}, P., {Lin}, D.~N.~C., \& {Mardling}, R.~A. 2001, \apj, 548, 466

\bibitem[{{Boltzmann}(1884)}]{1884AnP...258..291B}
{Boltzmann}, L. 1884, Annalen der Physik, 258, 291

\bibitem[{{Brooker} \& {Olle}(1955)}]{1955MNRAS.115..101B}
{Brooker}, R.~A. \& {Olle}, T.~W. 1955, \mnras, 115, 101

\bibitem[{{{\c C}ak{\i}rl{\i}} {et~al.}(2009){{\c C}ak{\i}rl{\i}}, {Ibanoglu},
  \& {G{\"u}ng{\"o}r}}]{2009NewA...14..496C}
{{\c C}ak{\i}rl{\i}}, {\"O}., {Ibanoglu}, C., \& {G{\"u}ng{\"o}r}, C. 2009, New
  Astronomy, 14, 496

\bibitem[{{Chabrier} \& {Baraffe}(2000)}]{2000ARA&A..38..337C}
{Chabrier}, G. \& {Baraffe}, I. 2000, \araa, 38, 337

\bibitem[{{Chabrier} {et~al.}(2000){Chabrier}, {Baraffe}, {Allard}, \&
  {Hauschildt}}]{2000ApJ...542..464C}
{Chabrier}, G., {Baraffe}, I., {Allard}, F., \& {Hauschildt}, P. 2000, \apj,
  542, 464

\bibitem[{{Chabrier} {et~al.}(2007){Chabrier}, {Gallardo}, \&
  {Baraffe}}]{2007A&A...472L..17C}
{Chabrier}, G., {Gallardo}, J., \& {Baraffe}, I. 2007, \aap, 472, L17

\bibitem[{{Coughlin} \& {Shaw}(2007)}]{2007JSARA...1....7C}
{Coughlin}, J.~L. \& {Shaw}, J.~S. 2007, Journal of the Southeastern
  Association for Research in Astronomy, 1, 7

\bibitem[{{D'Antona} \& {Mazzitelli}(1997)}]{1997MmSAI..68..807D}
{D'Antona}, F. \& {Mazzitelli}, I. 1997, Memorie della Societa Astronomica
  Italiana, 68, 807

\bibitem[{{Dreizler} {et~al.}(2009){Dreizler}, {Reiners}, {Homeier}, \&
  {Noll}}]{2009A&A...499..615D}
{Dreizler}, S., {Reiners}, A., {Homeier}, D., \& {Noll}, M. 2009, \aap, 499,
  615

\bibitem[{{Ferraz-Mello} {et~al.}(2008){Ferraz-Mello}, {Rodr{\'{\i}}guez}, \&
  {Hussmann}}]{2008CeMDA.101..171F}
{Ferraz-Mello}, S., {Rodr{\'{\i}}guez}, A., \& {Hussmann}, H. 2008, Celestial
  Mechanics and Dynamical Astronomy, 101, 171, (FM08)

\bibitem[{{Gavrilov} \& {Zharkov}(1977)}]{1977Icar...32..443G}
{Gavrilov}, S.~V. \& {Zharkov}, V.~N. 1977, Icarus, 32, 443

\bibitem[{{Goldreich} \& {Soter}(1966)}]{1966Icar....5..375G}
{Goldreich}, P. \& {Soter}, S. 1966, Icarus, 5, 375

\bibitem[{{G{\'o}mez Maqueo Chew} {et~al.}(2009){G{\'o}mez Maqueo Chew},
  {Stassun}, {Pr{\v s}a}, \& {Mathieu}}]{2009ApJ...699.1196G}
{G{\'o}mez Maqueo Chew}, Y., {Stassun}, K.~G., {Pr{\v s}a}, A., \& {Mathieu},
  R.~D. 2009, \apj, 699, 1196

\bibitem[{{Greenberg}(2009)}]{2009ApJ...698L..42G}
{Greenberg}, R. 2009, \apjl, 698, L42

\bibitem[{{Greenberg} {et~al.}(2008){Greenberg}, {Barnes}, \&
  {Jackson}}]{2008DPS....40.0403G}
{Greenberg}, R., {Barnes}, R., \& {Jackson}, B. 2008, in Bulletin of the
  American Astronomical Society, Vol.~40, Bulletin of the American Astronomical
  Society, 391--+

\bibitem[{{Gu} {et~al.}(2003){Gu}, {Lin}, \&
  {Bodenheimer}}]{2003ApJ...588..509G}
{Gu}, P., {Lin}, D.~N.~C., \& {Bodenheimer}, P.~H. 2003, \apj, 588, 509

\bibitem[{{Guenther} {et~al.}(2001){Guenther}, {Torres}, {Batalha}, {Joergens},
  {Neuh{\"a}user}, {Vijapurkar}, \& {Mundt}}]{2001A&A...366..965G}
{Guenther}, E.~W., {Torres}, G., {Batalha}, N., {et~al.} 2001, \aap, 366, 965

\bibitem[{{Hut}(1981)}]{1981A&A....99..126H}
{Hut}, P. 1981, \aap, 99, 126, (Hut81)

\bibitem[{{Ibgui} \& {Burrows}(2009)}]{2009ApJ...700.1921I}
{Ibgui}, L. \& {Burrows}, A. 2009, \apj, 700, 1921

\bibitem[{{Ibgui} {et~al.}(2009{\natexlab{a}}){Ibgui}, {Burrows}, \&
  {Spiegel}}]{2009arXiv0910.4394I}
{Ibgui}, L., {Burrows}, A., \& {Spiegel}, D.~S. 2009{\natexlab{a}}, ArXiv
  e-prints

\bibitem[{{Ibgui} {et~al.}(2009{\natexlab{b}}){Ibgui}, {Spiegel}, \&
  {Burrows}}]{2009arXiv0910.5928I}
{Ibgui}, L., {Spiegel}, D.~S., \& {Burrows}, A. 2009{\natexlab{b}}, ArXiv
  e-prints

\bibitem[{{Ioannou} \& {Lindzen}(1993)}]{1993ApJ...406..266I}
{Ioannou}, P.~J. \& {Lindzen}, R.~S. 1993, \apj, 406, 266

\bibitem[{{Jackson} {et~al.}(2008{\natexlab{a}}){Jackson}, {Barnes}, \&
  {Greenberg}}]{2008MNRAS.391..237J}
{Jackson}, B., {Barnes}, R., \& {Greenberg}, R. 2008{\natexlab{a}}, \mnras,
  391, 237

\bibitem[{{Jackson} {et~al.}(2008{\natexlab{b}}){Jackson}, {Greenberg}, \&
  {Barnes}}]{2008ApJ...681.1631J}
{Jackson}, B., {Greenberg}, R., \& {Barnes}, R. 2008{\natexlab{b}}, \apj, 681,
  1631

\bibitem[{{Kippenhahn} \& {Weigert}(1990)}]{1990sse..book.....K}
{Kippenhahn}, R. \& {Weigert}, A. 1990, {Stellar Structure and Evolution}
  (Stellar Structure and Evolution, XVI, 468 pp.~192 figs..~ Springer-Verlag
  Berlin Heidelberg New York.~Also Astronomy and Astrophysics Library)

\bibitem[{{Levrard} {et~al.}(2007){Levrard}, {Correia}, {Chabrier}, {Baraffe},
  {Selsis}, \& {Laskar}}]{2007A&A...462L...5L}
{Levrard}, B., {Correia}, A.~C.~M., {Chabrier}, G., {et~al.} 2007, \aap, 462,
  L5, (Lev07)

\bibitem[{{Mardling} \& {Lin}(2002)}]{2002ApJ...573..829M}
{Mardling}, R.~A. \& {Lin}, D.~N.~C. 2002, \apj, 573, 829

\bibitem[{{Mardling} \& {Lin}(2004)}]{2004ApJ...614..955M}
{Mardling}, R.~A. \& {Lin}, D.~N.~C. 2004, \apj, 614, 955

\bibitem[{{Marley} {et~al.}(2007){Marley}, {Fortney}, {Hubickyj},
  {Bodenheimer}, \& {Lissauer}}]{2007ApJ...655..541M}
{Marley}, M.~S., {Fortney}, J.~J., {Hubickyj}, O., {Bodenheimer}, P., \&
  {Lissauer}, J.~J. 2007, \apj, 655, 541

\bibitem[{{McLaughlin}(1924)}]{1924ApJ....60...22M}
{McLaughlin}, D.~B. 1924, \apj, 60, 22

\bibitem[{{Miller} {et~al.}(2009){Miller}, {Fortney}, \&
  {Jackson}}]{2009ApJ...702.1413M}
{Miller}, N., {Fortney}, J.~J., \& {Jackson}, B. 2009, \apj, 702, 1413

\bibitem[{{Mohanty} {et~al.}(2007){Mohanty}, {Baraffe}, \&
  {Chabrier}}]{2007IAUS..239..197M}
{Mohanty}, S., {Baraffe}, I., \& {Chabrier}, G. 2007, in IAU Symposium, Vol.
  239, IAU Symposium, ed. F.~{Kupka}, I.~{Roxburgh}, \& K.~{Chan}, 197--204

\bibitem[{{Mohanty} {et~al.}(2009){Mohanty}, {Stassun}, \&
  {Mathieu}}]{2009ApJ...697..713M}
{Mohanty}, S., {Stassun}, K.~G., \& {Mathieu}, R.~D. 2009, \apj, 697, 713

\bibitem[{{Morales} {et~al.}(2009){Morales}, {Ribas}, {Jordi}, {Torres},
  {Gallardo}, {Guinan}, {Charbonneau}, {Wolf}, {Latham}, {Anglada-Escud{\'e}},
  {Bradstreet}, {Everett}, {O'Donovan}, {Mandushev}, \&
  {Mathieu}}]{2009ApJ...691.1400M}
{Morales}, J.~C., {Ribas}, I., {Jordi}, C., {et~al.} 2009, \apj, 691, 1400

\bibitem[{{Neron de Surgy} \& {Laskar}(1997)}]{1997A&A...318..975N}
{Neron de Surgy}, O. \& {Laskar}, J. 1997, \aap, 318, 975

\bibitem[{{Ogilvie} \& {Lin}(2004)}]{2004ApJ...610..477O}
{Ogilvie}, G.~I. \& {Lin}, D.~N.~C. 2004, \apj, 610, 477

\bibitem[{{Ohta} {et~al.}(2005){Ohta}, {Taruya}, \&
  {Suto}}]{2005ApJ...622.1118O}
{Ohta}, Y., {Taruya}, A., \& {Suto}, Y. 2005, \apj, 622, 1118

\bibitem[{{Peale} \& {Greenberg}(1980)}]{1980LPI....11..871P}
{Peale}, S.~J. \& {Greenberg}, R.~J. 1980, in Lunar and Planetary Institute
  Conference Abstracts, Vol.~11, Lunar and Planetary Institute Conference
  Abstracts, 871--873

\bibitem[{{Ray} {et~al.}(2001){Ray}, {Eanes}, \&
  {Lemoine}}]{2001GeoJI.144..471R}
{Ray}, R.~D., {Eanes}, R.~J., \& {Lemoine}, F.~G. 2001, Geophysical Journal
  International, 144, 471

\bibitem[{{Reiners} {et~al.}(2007){Reiners}, {Seifahrt}, {Stassun}, {Melo}, \&
  {Mathieu}}]{2007ApJ...671L.149R}
{Reiners}, A., {Seifahrt}, A., {Stassun}, K.~G., {Melo}, C., \& {Mathieu},
  R.~D. 2007, \apjl, 671, L149

\bibitem[{{Ribas} {et~al.}(2008){Ribas}, {Morales}, {Jordi}, {Baraffe},
  {Chabrier}, \& {Gallardo}}]{2008MmSAI..79..562R}
{Ribas}, I., {Morales}, J.~C., {Jordi}, C., {et~al.} 2008, Memorie della
  Societa Astronomica Italiana, 79, 562

\bibitem[{{Rossiter}(1924)}]{1924ApJ....60...15R}
{Rossiter}, R.~A. 1924, \apj, 60, 15

\bibitem[{{Scholz} \& {Eisl{\"o}ffel}(2005)}]{2005A&A...429.1007S}
{Scholz}, A. \& {Eisl{\"o}ffel}, J. 2005, \aap, 429, 1007

\bibitem[{{Stassun} {et~al.}(2009){Stassun}, {Hebb}, {L{\'o}pez-Morales}, \&
  {Pr{\v s}a}}]{2009IAUS..258..161S}
{Stassun}, K.~G., {Hebb}, L., {L{\'o}pez-Morales}, M., \& {Pr{\v s}a}, A. 2009,
  in IAU Symposium, Vol. 258, IAU Symposium, ed. E.~E. {Mamajek}, D.~R.
  {Soderblom}, \& R.~F.~G. {Wyse}, 161--170

\bibitem[{{Stassun} {et~al.}(2008){Stassun}, {Mathieu}, {Cargile}, {Aarnio},
  {Stempels}, \& {Geller}}]{2008Natur.453.1079S}
{Stassun}, K.~G., {Mathieu}, R.~D., {Cargile}, P.~A., {et~al.} 2008, \nat, 453,
  1079

\bibitem[{{Stassun} {et~al.}(2006){Stassun}, {Mathieu}, \&
  {Valenti}}]{2006Natur.440..311S}
{Stassun}, K.~G., {Mathieu}, R.~D., \& {Valenti}, J.~A. 2006, \nat, 440, 311

\bibitem[{{Stassun} {et~al.}(2007){Stassun}, {Mathieu}, \&
  {Valenti}}]{2007ApJ...664.1154S}
{Stassun}, K.~G., {Mathieu}, R.~D., \& {Valenti}, J.~A. 2007, \apj, 664, 1154

\bibitem[{{Stefan}(1879)}]{Stefan}
{Stefan}, J. 1879, Sitzungsberichte der mathematisch-naturwissenschaftlichen
  Classe der kaiserlichen Akademie der Wissenschaften, 79, 391

\bibitem[{{Winn} {et~al.}(2005){Winn}, {Noyes}, {Holman}, {Charbonneau},
  {Ohta}, {Taruya}, {Suto}, {Narita}, {Turner}, {Johnson}, {Marcy}, {Butler},
  \& {Vogt}}]{2005ApJ...631.1215W}
{Winn}, J.~N., {Noyes}, R.~W., {Holman}, M.~J., {et~al.} 2005, \apj, 631, 1215

\bibitem[{{Wisdom}(2008)}]{2008Icar..193..637W}
{Wisdom}, J. 2008, Icarus, 193, 637, (Wis08)

\bibitem[{{Wuchterl}(2005)}]{2005AN....326..905W}
{Wuchterl}, G. 2005, Astronomische Nachrichten, 326, 905

\bibitem[{{Yoder}(1979)}]{1979Natur.279..767Y}
{Yoder}, C.~F. 1979, \nat, 279, 767

\bibitem[{{Zhang} \& {Hamilton}(2008)}]{2008Icar..193..267Z}
{Zhang}, K. \& {Hamilton}, D.~P. 2008, Icarus, 193, 267

\end{thebibliography}

\clearpage

\begin{landscape}
\pagestyle{empty}
\topmargin 4cm 

\begin{figure*}
  \centering
  \scalebox{0.52}{\includegraphics{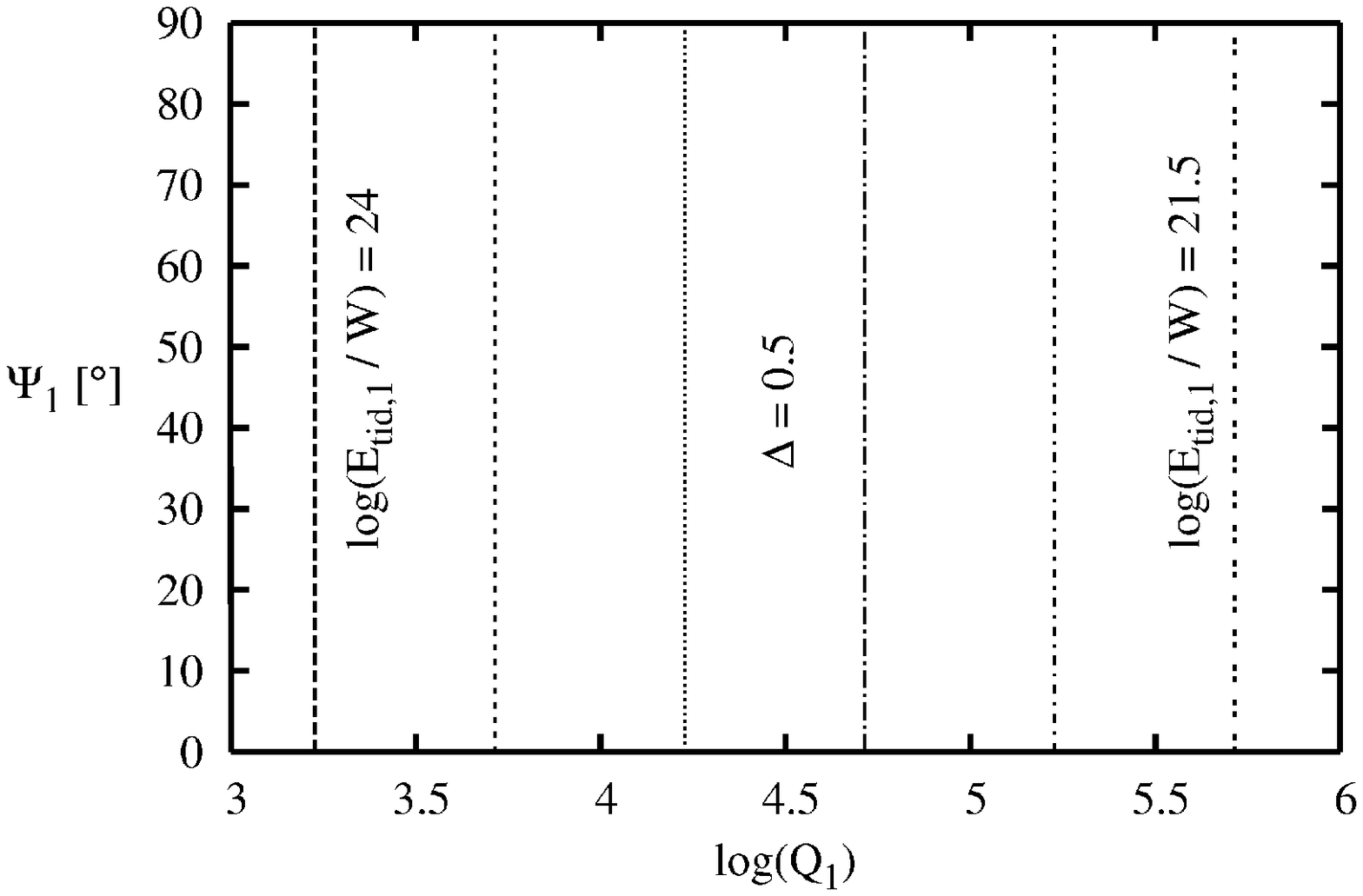}}
  \hspace{1cm}
  \scalebox{0.52}{\includegraphics{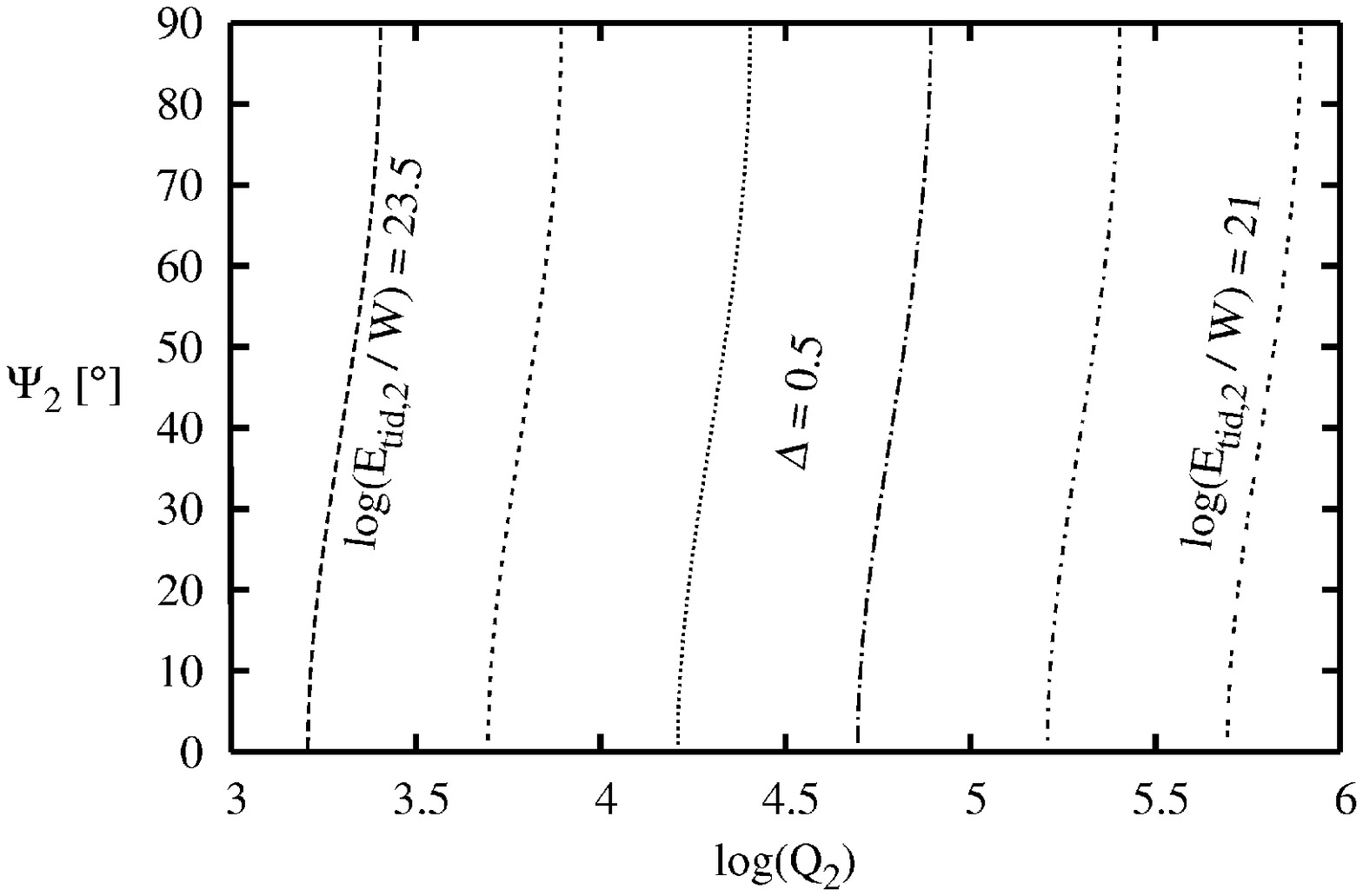}}
  \caption{Tidal heating after model \#1. \textit{Left}: (Primary) Projection of $\dot{E}_{\mathrm{tid},1}^{\mathrm{\#1}}$ onto the $\log(Q_1)$-$\psi_1$ plane. The stepsize between contour lines is chosen to be $\Delta = 0.5$ in $\log(\dot{E}_{\mathrm{tid},1}^{\mathrm{\#1}}/\mathrm{W})$. \textit{Right}: (Secondary) Projection of $\dot{E}_{\mathrm{tid},2}^{\mathrm{\#1}}$ onto the $\log(Q_2)$-$\psi_2$ plane. Although there is a dependence on $\psi_2$, the energy rates at a fixed value for the quality factor are smaller than those for the primary.}
\label{fig:E_mod1}
\end{figure*}

\begin{figure*}
  \centering
  \scalebox{0.45}{\includegraphics{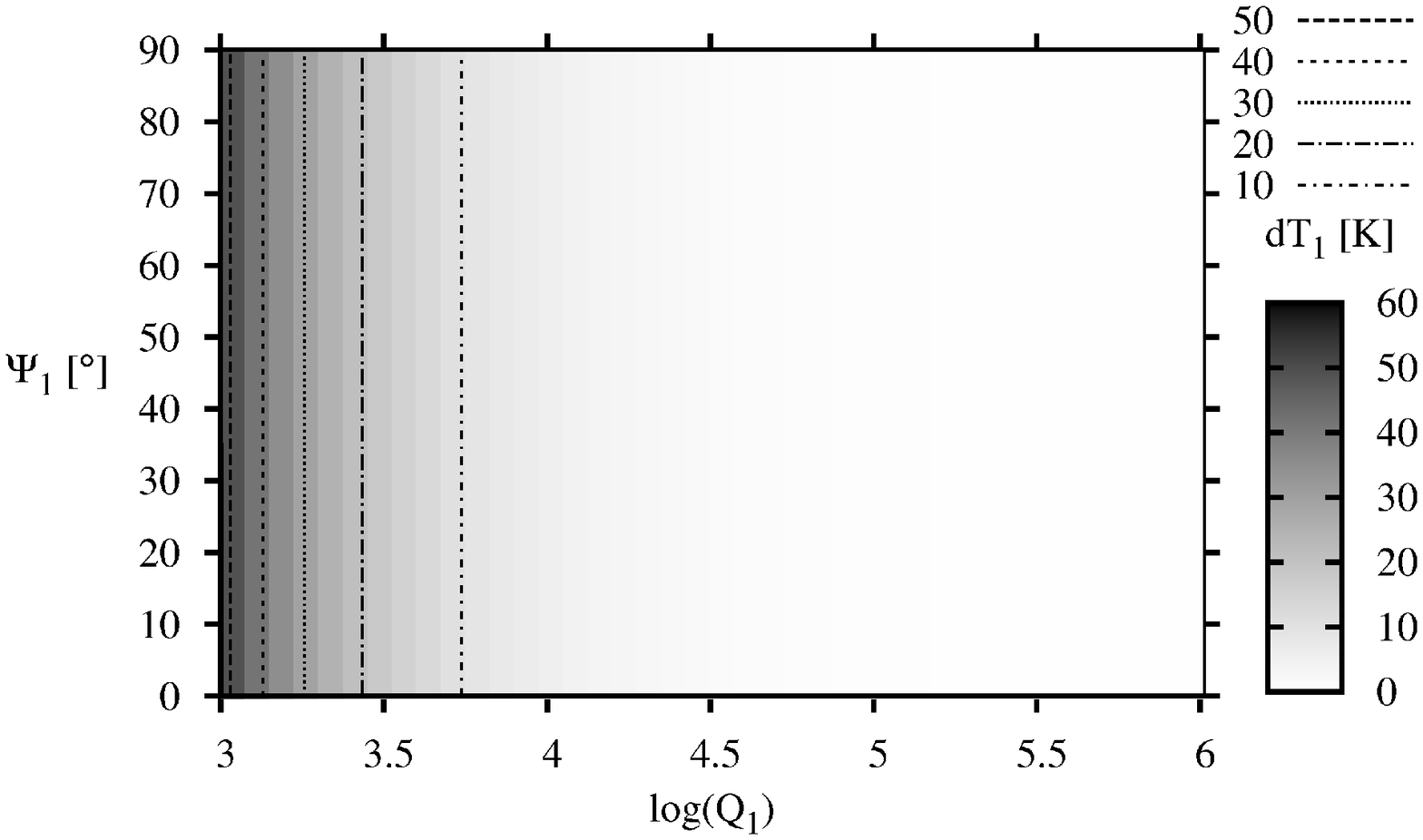}}
  \hspace{0.5cm}
  \scalebox{0.45}{\includegraphics{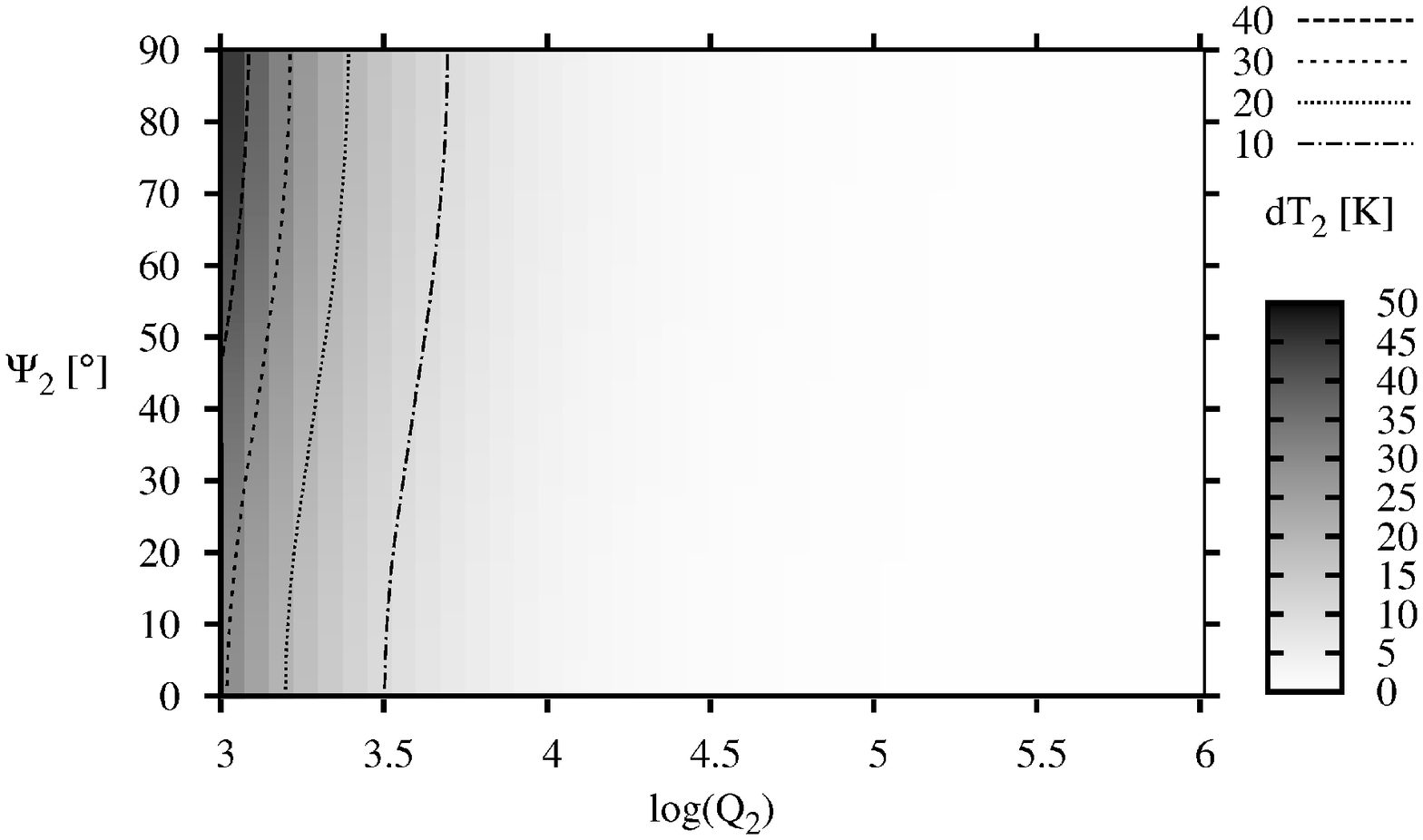}}
  \caption{Temperature increase after model \#1. \textit{Left}: (Primary) Projection of $d$T$_1$ onto the $\log(Q_1)$-$\psi_1$ plane. For a significant temperature increase, $Q_1$ would have to be much smaller than $10^{3.5}$, but such a temperature increase is not observed in the primary. \textit{Right}: (Secondary) Projection of $d$T$_2$ onto the $\log(Q_2)$-$\psi_2$ plane. Even for very low values of $Q_2$ and high obliquities $\psi_2$ the observed temperature increase cannot be reconstructed. For any given point in the $\psi$-$\log(Q)$ plane, $\mathrm{d}T_2 < \mathrm{d}T_1$, which does not support the observed temperature reversal.}
\label{fig:dT_mod1}
\end{figure*}

\end{landscape}

\begin{landscape}
\pagestyle{empty}
\topmargin 4cm 

\begin{figure*}
  \centering
  \scalebox{0.52}{\includegraphics{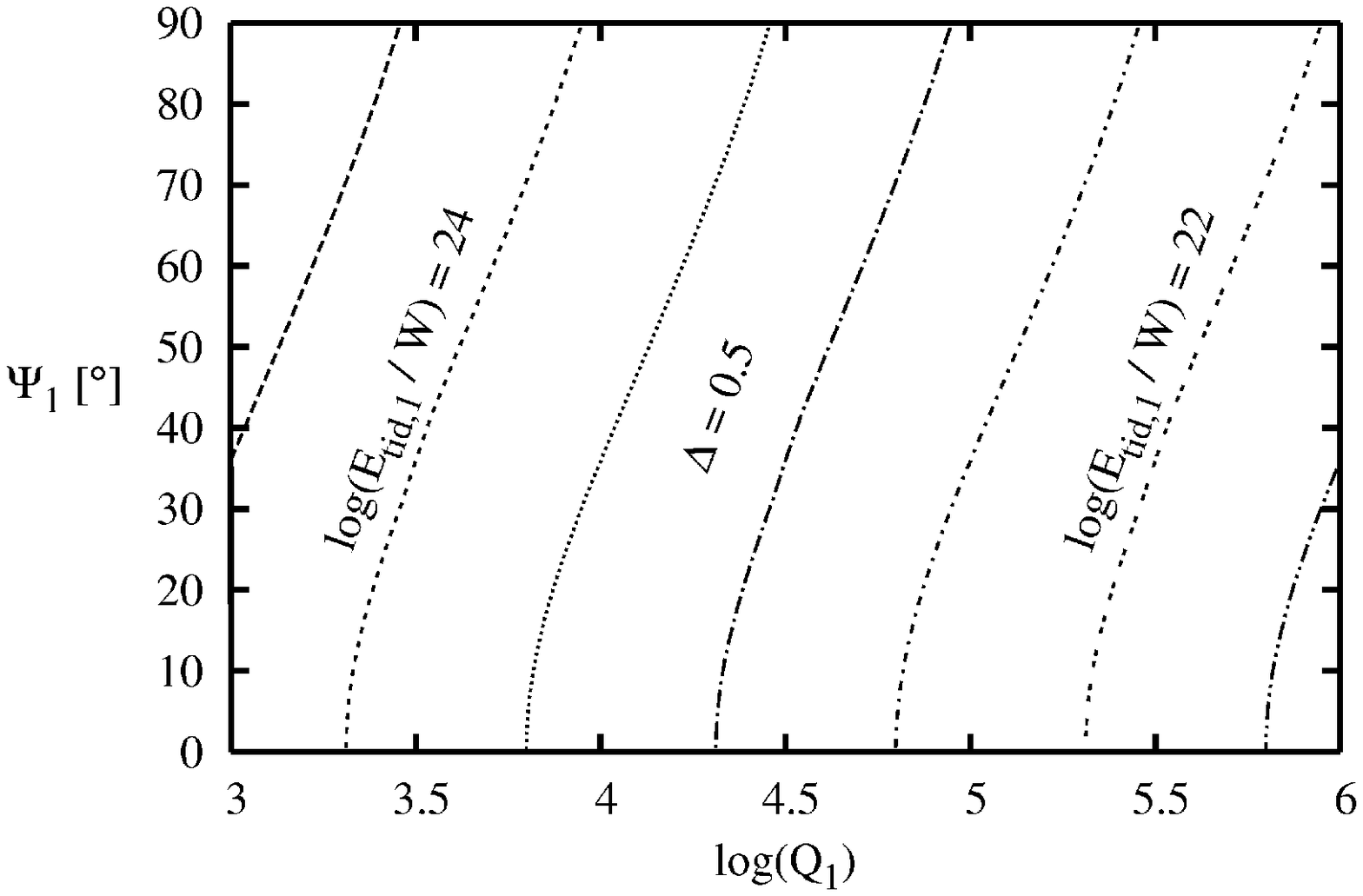}}
  \hspace{1cm}
  \scalebox{0.52}{\includegraphics{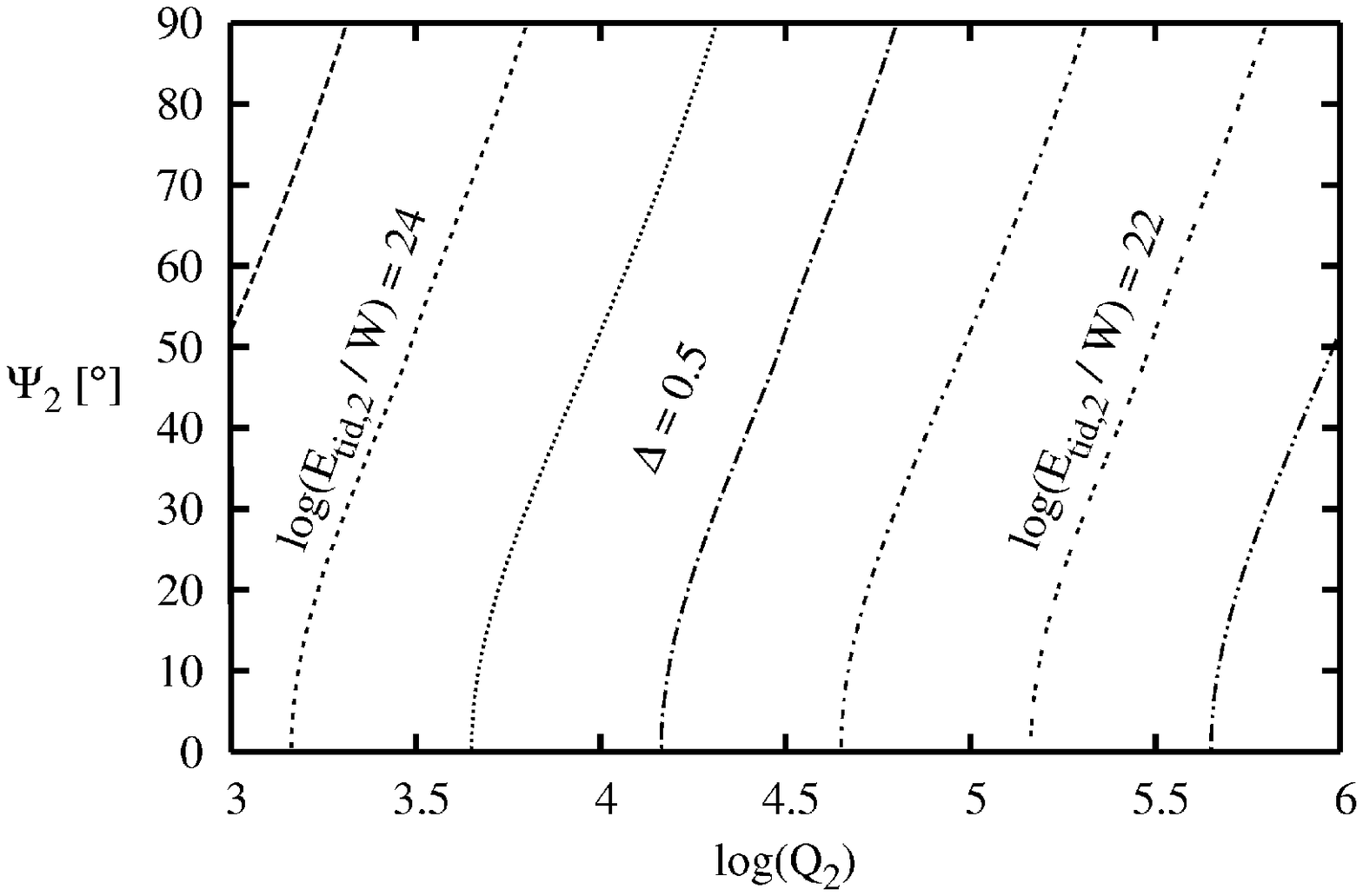}}
  \caption{Tidal heating after model \#2.  \textit{Left}: (Primary) Projection of $\dot{E}_{\mathrm{tid},1}^{\mathrm{\#2}}$ onto the $\log(Q_1)$-$\psi_1$ plane. The stepsize between contour lines is chosen to be $\Delta = 0.5$ in $\log(\dot{E}_{\mathrm{tid},1}^{\mathrm{\#2}}/\mathrm{W})$. The tidal energy rates strongly depend on a putative obliquity, different from model \#1. \textit{Right}: (Secondary) Projection of $\dot{E}_{\mathrm{tid},2}^{\mathrm{\#2}}$ onto the $\log(Q_2)$-$\psi_2$ plane. For the three models (\#1, \#2, and \#4) invoking $Q$ and $\psi$, these rates are the highest of all for any given point in the $\psi$-$\log(Q)$ plane -- for the primary as well as for the secondary.}
  \label{fig:E_mod2}
\end{figure*}

\begin{figure*}
  \centering
  \scalebox{0.45}{\includegraphics{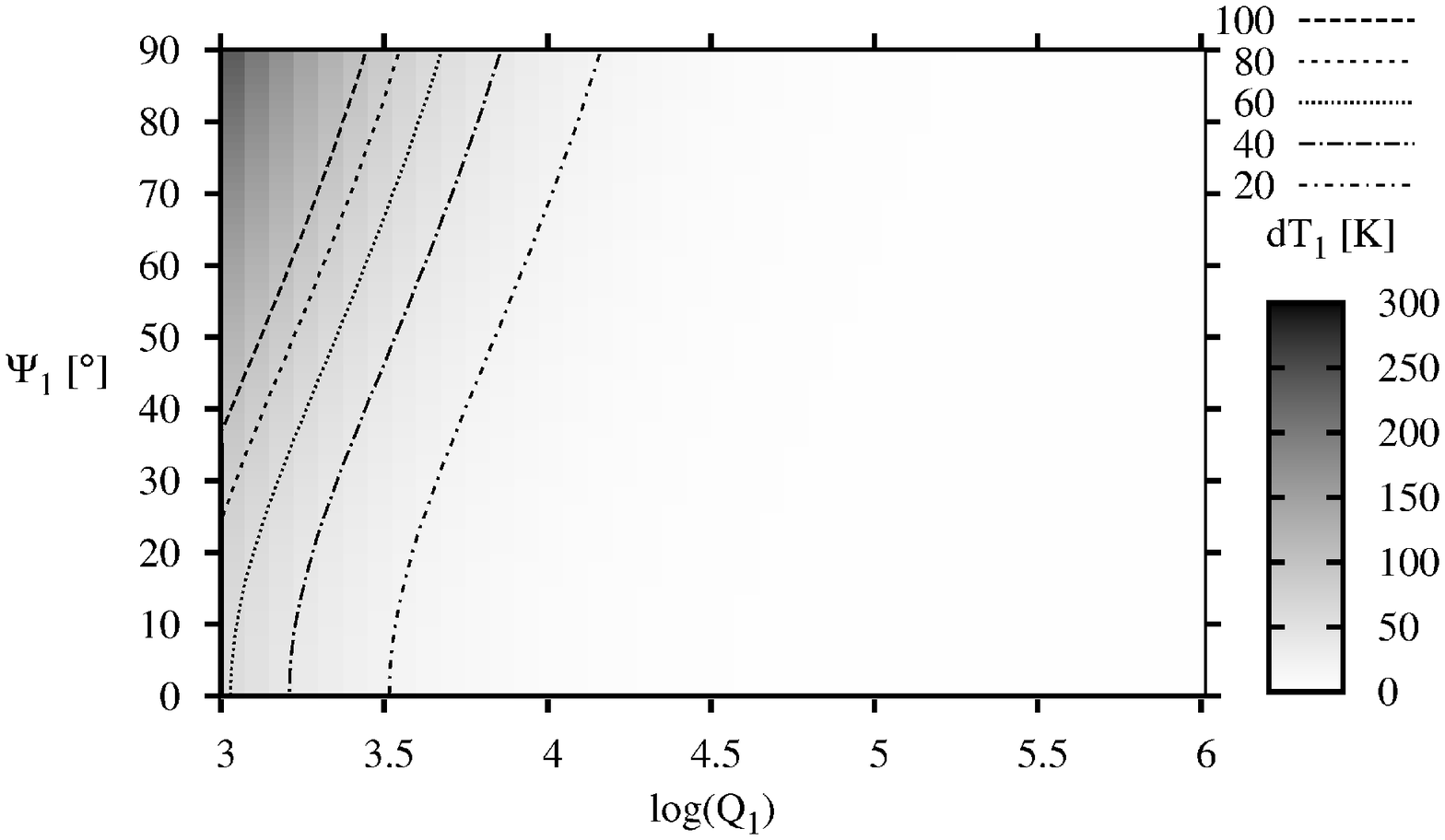}}
  \hspace{0.5cm}
  \scalebox{0.45}{\includegraphics{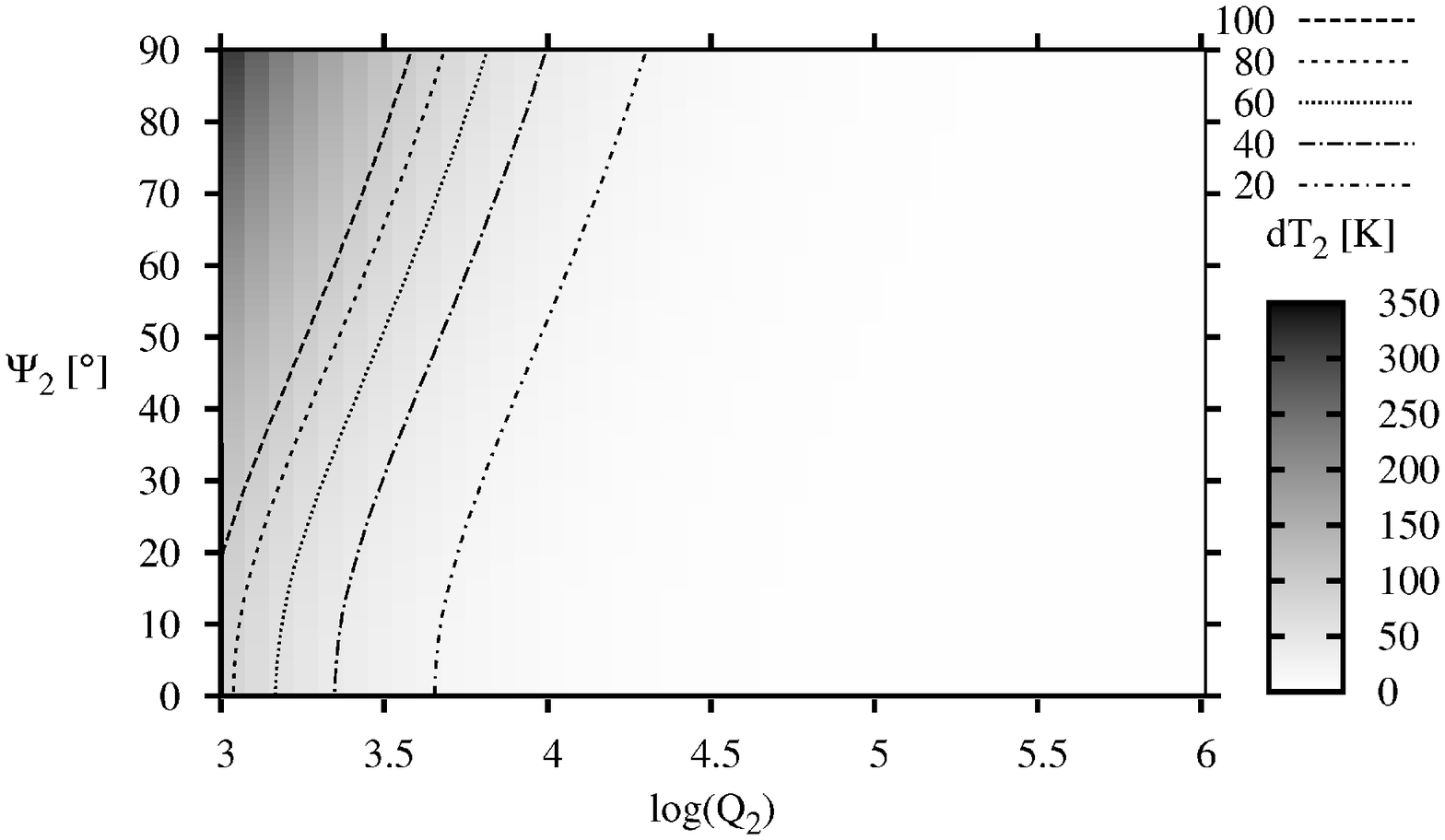}}
  \caption{Temperature increase after model \#2. \textit{Left}: (Primary) Projection of $d$T$_1$ onto the $\log(Q_1)$-$\psi_1$ plane. \textit{Right}: (Secondary) Projection of $d$T$_2$ onto the $\log(Q_2)$-$\psi_2$ plane. For any given location in the $\log(Q)$-$\psi$ plane, model \#2 yields the strongest temperature increase compared to the other models -- both for the primary and the secondary, respectively. For a given spot in $Q$-$\psi$ space there is an inversion in temperature increase: $\mathrm{d}T_2 > \mathrm{d}T_1$, i.e. the less massive BD is heated more.}
  \label{fig:dT_mod2}
\end{figure*}
\end{landscape}

\begin{landscape}
\pagestyle{empty}
\topmargin 4cm 

\begin{figure*}
  \centering
  \scalebox{0.52}{\includegraphics{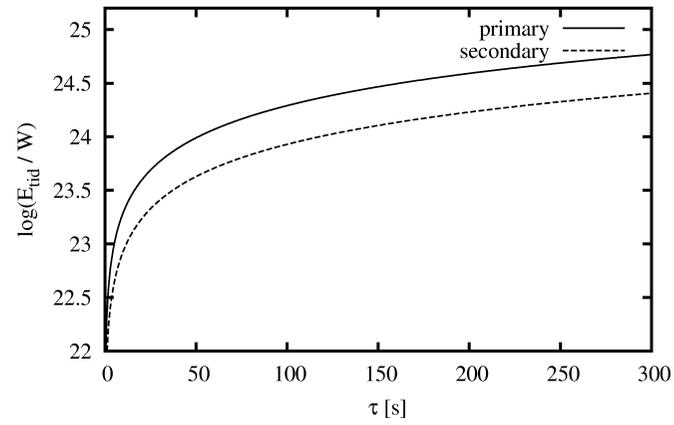}}
  \caption{Tidal heating within the primary (solid line) and secondary (dashed line) after model \#3. While the tidal heating rate of the secondary becomes comparable to its observed luminosity overshoot for $\tau \gtrsim 100$\,s, if the same $\tau$ is applied to the primary, heating within the primary would lead to a larger luminosity than is observed.}
  \label{fig:E_mod3}
\end{figure*}

\begin{figure*}
  \centering
  \scalebox{0.52}{\includegraphics{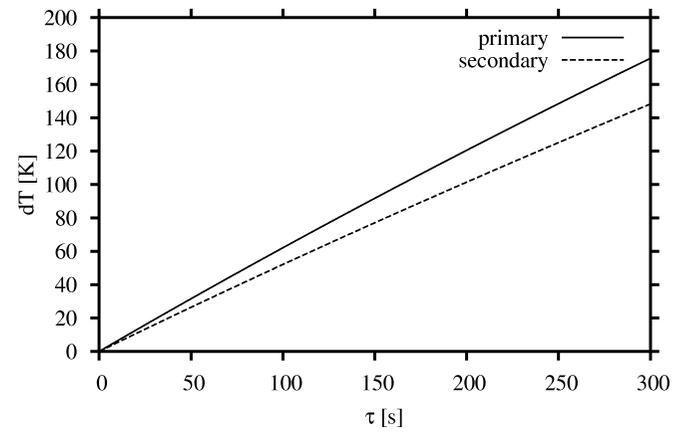}}
  \caption{Temperature increase of the primary (solid line) and secondary (dashed line) after model \#3. Contrary to what is observed, the primary would be hotter than the secondary.}
  \label{fig:dT_mod3}
\end{figure*}

\end{landscape}

\begin{landscape}
\pagestyle{empty}
\topmargin 4cm 

\begin{figure*}
  \scalebox{0.52}{\includegraphics{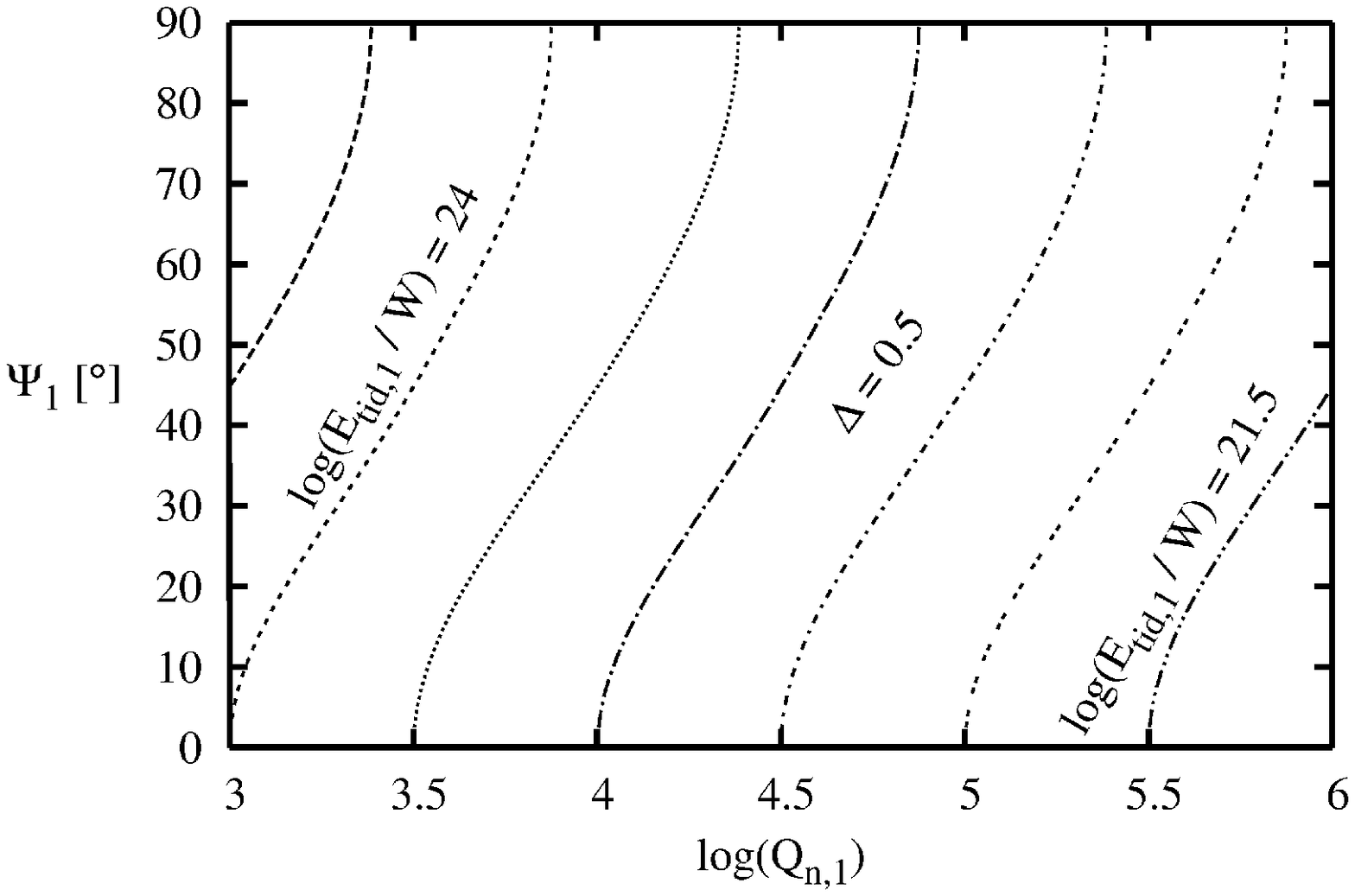}}
  \hspace{1.cm}
  \scalebox{0.52}{\includegraphics{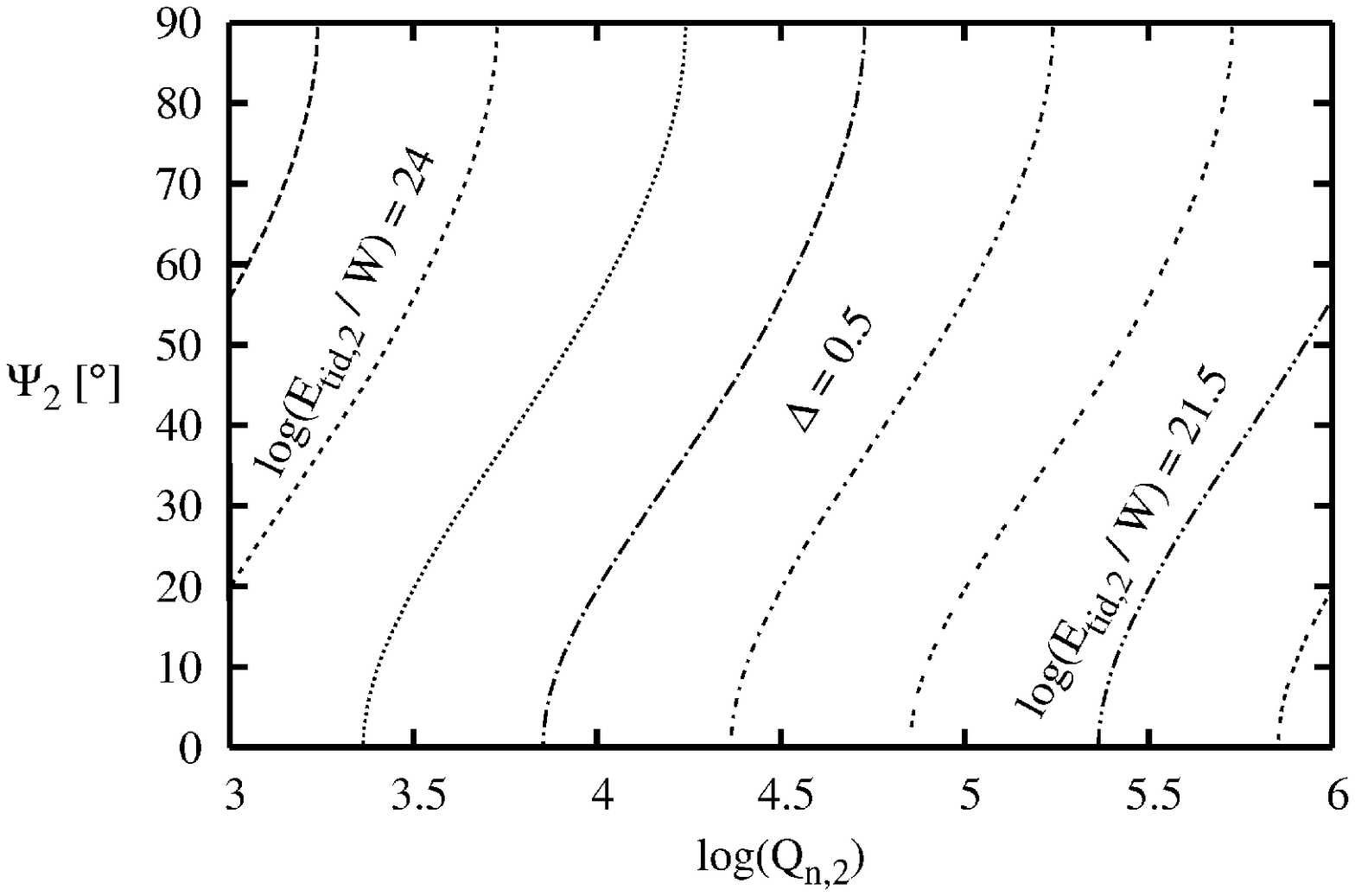}}
  \caption{Tidal heating after model \#4. \textit{Left}: (Primary) Projection of $\dot{E}_{\mathrm{tid},1}^{\mathrm{\#4}}$ onto the $\log(Q_1)$-$\psi_1$ plane. The stepsize between contour lines is chosen to be $\Delta = 0.5$ in $\log(\dot{E}_{\mathrm{tid},1}^{\mathrm{\#4}}/\mathrm{W})$. \textit{Right}: (Secondary) Projection of $\dot{E}_{\mathrm{tid},2}^{\mathrm{\#4}}$ onto the $\log(Q_2)$-$\psi_2$ plane.}
  \label{fig:E_mod4}
\end{figure*}

\begin{figure*}
  \scalebox{0.45}{\includegraphics{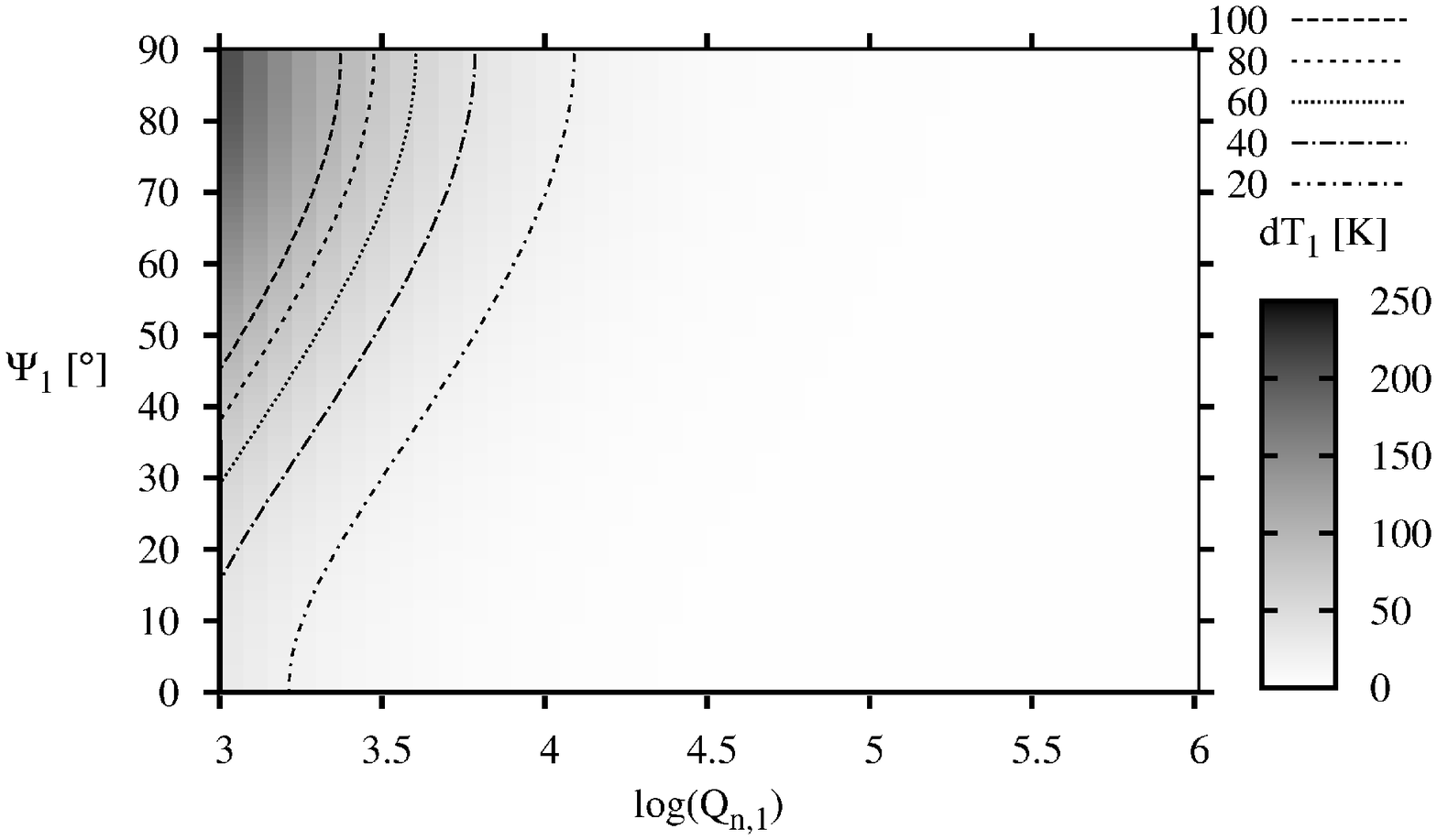}}
  \hspace{0.5cm}
  \scalebox{0.45}{\includegraphics{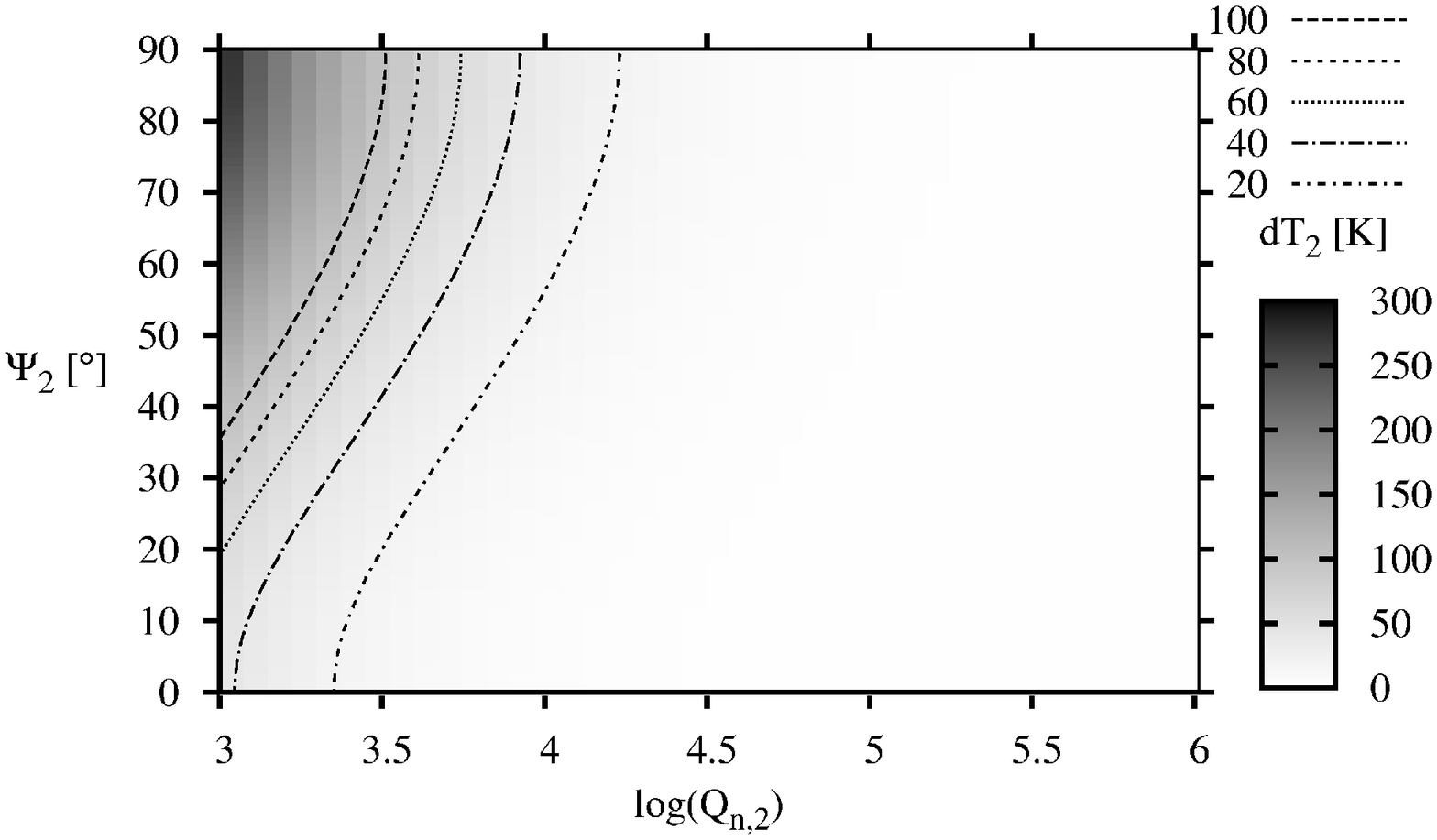}}
  \caption{Temperature increase after model \#4. \textit{Left}: (Primary) Projection of $d$T$_1$ onto the $\log(Q_1)$-$\psi_1$ plane. \textit{Right}: (Secondary) Projection of $d$T$_2$ onto the $\log(Q_2)$-$\psi_2$ plane. For $\log(Q_2) \approx 3.5$ and an obliquity of $\psi_2 \approx 70^{\circ}$ the temperature increase of the secondary becomes similar to the observed one. For the whole range of $Q$ and $\psi$ there is an inversion in temperature increase, similar to model \#2: $\mathrm{d}T_2 > \mathrm{d}T_1$.}
  \label{fig:dT_mod4}
\end{figure*}

\end{landscape}

\clearpage

\end{document}